\documentclass[12pt,preprint]{aastex}
\usepackage{graphicx}

\newcommand{\simgt}{\lower.5ex\hbox{$\;\buildrel>\over\sim\;$}}
\newcommand{\simlt}{\lower.5ex\hbox{$\;\buildrel<\over\sim\;$}}

\newcommand{\hii}{\rm H {\sc ii}}
\newcommand{\heii}{\rm He {\sc ii}}

\newcommand{\htwo}{\rm H$_2$}

\newcommand{\neii}{\rm [Ne\,{\sc ii}]}
\newcommand{\neiii}{\rm [Ne\,{\sc iii}]}
\newcommand{\arii}{\rm [Ar\,{\sc ii}]}
\newcommand{\ariii}{\rm [Ar\,{\sc iii}]}

\newcommand{\siii}{\rm [S\,{\sc iii}]}

\newcommand{\siv}{\rm [S\,{\sc iv}]}
\newcommand{\oiv}{\rm [O\,{\sc iv}]}
\newcommand{\fev}{\rm [Fe\,{\sc v}]}
\newcommand{\feii}{\rm [Fe\,{\sc ii}]}

\newcommand{\water}{\rm H$_2$O}
\newcommand{\jh}{{$J-H$}}
\newcommand{\hk}{{$H-K$}}
\newcommand{\jk}{{$J-K$}}

\newcommand{\hst}{{\sl HST}}

\newcommand{\hb}{\ensuremath{{\rm H}\beta}}

\newcommand{\mic}{$\mu${\rm m}}

\newcommand{\hh}{Haro\,3}
\newcommand{\mrk}{Mrk\,996}
\newcommand{\spitzer}{{\sl Spitzer}}

\newcommand{\mopex}{{\it MOPEX}}
\newcommand{\spice}{{\it SPICE}}

\shorttitle{The \spitzer\ View of \mrk}
\shortauthors{Thuan et al.}

\begin{document}

\title{The \spitzer\ View of Low-Metallicity Star Formation: II. \mrk, a 
Blue Compact Dwarf Galaxy with an Extremely Dense Nucleus}

\author{
	Trinh~X.~Thuan\altaffilmark{1}, 
	Leslie~K.~Hunt\altaffilmark{2} and
	Yuri~I.~Izotov\altaffilmark{3} 
}
\altaffiltext{1}{Astronomy Department, University of Virginia, P.O. Box 400325,
Charlottesville, VA 22904-4325, USA: txt@virginia.edu}
\altaffiltext{2}{INAF-Istituto di Radioastronomia-Sez.\ Firenze, 
L.go Fermi 5, I-50125 Firenze, Italy; hunt@arcetri.astro.it.}
\altaffiltext{3}{Main Astronomical Observatory, National Academy of Sciences of Ukraine, 
03680 Kiev, Ukraine; izotov@mao.kiev.ua}

\begin{abstract}
We present new \spitzer, UKIRT and MMT observations of the
blue compact dwarf galaxy (BCD) \mrk, with an oxygen abundance
of 12$+$log(O/H)\,=\,8.0. This galaxy has the peculiarity of possessing an 
extraordinarily dense nuclear star-forming region, with a central density of 
$\sim$ 10$^6$ cm$^{-3}$. The nuclear region of \mrk\ is characterized by 
several unusual properties: 
a very red color \jk\ = 1.8, broad and narrow emission-line components, and ionizing radiation
as hard as 54.9 eV, as implied by the presence of the \oiv\ 25.89\,\micron\ 
line.
The nucleus is located within an exponential disk 
with colors consistent with a single stellar population of age $\ga$ 1 Gyr. 
The infrared morphology of \mrk\ changes with wavelength.
IRAC 4.5\,\micron\ images show extended stellar photospheric emission from the 
body of the galaxy, and an extremely red nuclear point source, indicative of hot dust; 
IRAC 8\,\micron\ images show extended PAH emission from the surrounding ISM
and a bright nucleus; 
MIPS 24 and 70 images consist of bright point sources associated with the warm
nuclear dust; and 160 \,\micron\ images map the cooler extended dust 
associated with older stellar populations.
The IRS spectrum shows strong narrow 
Polycyclic Aromatic Hydrocarbon (PAH) emission, with 
narrow line widths and equivalent widths that are high for the metallicity of 
\mrk. 
Gaseous nebular fine-structure lines are also seen.
A CLOUDY model which accounts for both the optical and mid-infrared (MIR)
lines requires that 
they originate in two distinct \hii\ regions: a very dense 
\hii\ region of radius $\sim$ 580 pc 
with densities declining from $\sim$ 10$^6$ at the center 
to a few hundreds cm$^{-3}$ at the outer radius, where most of
the optical lines arise; and a \hii\ region with a density of 
$\sim$ 300  cm$^{-3}$ that is hidden in the optical but seen 
in the MIR. We suggest 
that the infrared lines arise mainly in the optically 
obscured \hii\ region
while they are strongly suppressed by collisional deexcitation 
in the optically visible one. The hard ionizing radiation needed to account for 
 the \oiv\ 25.89\,\micron\ line is most likely due to fast radiative 
shocks propagating in an 
interstellar medium. A hidden population of 
Wolf-Rayet stars of type WNE-w or a hidden AGN as sources of hard
 ionizing radiation are less likely possibilities.

\end{abstract}


\section{\label{sec:intro}Introduction}

Among blue compact dwarf galaxies (BCDs), the dwarf emission-line 
galaxy Mrk 996 ($M_B$\,=\,$-16.9$) occupies a place apart because of 
the extreme electron 
density at the center of its star-forming region.
{\sl HST} $V$ and $I$ images show that the bulk of 
the star formation occurs in a compact roughly circular 
high surface brightness  
nuclear region  of radius $\sim$ 340 pc, with evident dust patches to the 
north of it \citep{thuan96}. The nucleus (n) is located  
within an elliptically (E) shaped low surface brightness (LSB) 
component,
so that Mrk 996 belongs to the relatively rare class of nE BCDs 
\citep{loose85}. The extended envelope shows a distinct asymmetry, 
being more extended to the 
northeast side than to the southwest side, perhaps the sign of a past merger.
This asymmetry is also seen in the spatial distribution of the 
globular clusters 
around Mrk 996, these being seen mainly to the south of the galaxy.  

Mrk 996 has a heliocentric 
radial velocity of 1622 km s$^{-1}$ which gives it a distance of 21.7 Mpc, 
adopting a Hubble constant of 75 km s$^{-1}$ Mpc$^{-2}$ 
and including a very small correction for the Virgocentric flow \citep{thuanhi}.
At the adopted distance, 1 arcsec corresponds to a linear size of 105 pc.
\citet{thuan96} found the extended LSB component to possess 
an exponential disk structure with a small scale length of 0.42 kpc.
While Mrk 996 does not show an obvious spiral structure in the disk,
there is a spiral-like pattern in the nuclear star-forming region, which 
is no larger than 160 pc in radius. 
 
The UV and optical spectra of the nuclear star-forming region of Mrk 996
\citep{izotov92,thuan96} 
show remarkable features, suggesting very unusual physical conditions. The 
He {\sc i} line intensities are some 2 -- 4 times larger than those in 
normal BCDs. In the UV range, the N {\sc iii}] $\lambda$1750 and C {\sc iii}] $\lambda$1909 are particularly intense.
 Moreover, the line width depends on the degree of ionization 
of the ion. Thus low-ionization 
emission lines such as O$^+$, 
S$^+$ and N$^+$ have narrow widths, similar to those in 
other \hii\ regions, while high-ionization emission lines 
such as the helium lines, the O$^{++}$ and 
Ne$^{++}$ nebular lines and all auroral lines, show very broad line 
widths, $\ga$ 500 km s$^{-1}$. 
Such correlations of line widths with the degree of excitation
suggest different 
ionization zones with very distinct kinematical properties. 
\citet{thuan96} found that the usual one-zone low-density ionization-bounded 
\hii\ region model cannot be applied to the nuclear star-forming 
region of Mrk 996 without leading to unrealistic helium and 
heavy-element abundances. Instead, they showed that a two-zone 
density-bounded 
\hii\ region model, including an inner compact region with a  
central density of $\sim$ 10$^6$ cm$^{-3}$, some 4 orders of magnitude 
greater than the densities of normal \hii\ regions, 
together with an outer region with a lower density of $\sim$ 450 cm$^{-3}$, 
comparable to those of other \hii\ regions,
is needed to account for the 
observed line intensities. The large density gradient is probably caused by 
a mass outflow driven by the large population of Wolf-Rayet stars present in 
the galaxy. The gas outflow motions may account for the much broader 
line widths of the high-ionization lines originating in the 
dense inner region as compared to the low-ionization lines which 
originate in the less dense outer region. As for the high  
N {\sc iii}] $\lambda$1750, C {\sc iii}] $\lambda$1909 and He {\sc i} line 
intensities,
they can be understood by collisional excitation of these lines in the 
high density region.  
In the context of this model, the 
oxygen abundance of Mrk 996 is 12+log(O/H) = 8.0. Adopting 
12+log(O/H) = 8.65 for the Sun \citep{asplund05}, 
then Mrk 996 has a metallicity of 0.22 solar. 
Mrk 996 shows 
enhanced helium and nitrogen abundances, which can be accounted for by  
local pollution from Wolf-Rayet stars.
 
We present in this paper \spitzer\ \citep{werner04} mid-infrared (MIR) 
observations of Mrk 996. The extraordinary UV and optical properties 
of this galaxy, along with  
the evident presence of dust patches in the star-forming region 
of Mrk 996 on {\sl HST} optical images, make it a prime candidate for our 
Cycle 1 (PID 3139: P.I. Thuan) \spitzer\
observations. Our entire program consists of spectroscopic,
photometric and imaging observations of 23 BCDs with metallicities 
ranging from 1/20 to 1/2 that of the Sun, and its main aim is to
 study star formation in metal-poor environments 
and to understand  
how star formation and dust properties change as a function 
of metallicity and other physical parameters.
 This paper is the second of our \spitzer\ series, 
the first paper being 
on Haro 3, the most metal-rich BCD in our sample \citep{hunt06}. 
 
We present in \S2 our {\sl Spitzer} IRAC, MIPS and IRS  
observations of Mrk 996 and their 
reduction. We 
discuss in \S3 new complementary UKIRT near-infrared (NIR) imaging  
observations and optical MMT spectroscopic observations. 
In \S4, we discuss our imaging results: 
the IR morphology of the disk of old stars, and the 
nature of the very red, bright and dense nuclear IR source in Mrk 996. 
We also discuss 
the extended PAH emission. 
In \S5, we present our spectroscopic 
results: the PAH features and the IR fine-structure lines. In \S6, we
use the CLOUDY photoionization code \citep{cloudy,ferland98} 
to model the observed optical and IR emission-line 
intensities. We show that it is necessary to postulate two \hii\ regions:
one which is optically visible where the optical lines arise, and one 
which is optically hidden where the main part of the IR line emission 
arises.  We then  
discuss possible sources of hard ionizing radiation -- 
fast shocks, WNE-w stars or an AGN --  to account 
for the presence of 
the [O {\sc iv}] $\lambda$25.9$\mu$m, line. We  
summarize our conclusions in \S7.  

\section{Spitzer Observations}

In the context of our guest-observer program, we acquired
\spitzer\ observations of \mrk:  
IRAC images \citep{fazioirac} at 4.5 and 8 \micron, MIPS photometry
\citep{riekemips} at 24, 70 and 160 \micron\ and 
low- and high-resolution IRS spectra \citep{houckirs}. 
The data were obtained during the period from 16 December, 2004 to
3 January, 2005.
As the data 
reduction procedures have been described in detail in Paper I \citep{hunt06}, 
we give here only a summary.

\subsection{\label{sec:iracdata}IRAC Imaging} 

\begin{figure}
\vspace{0.2cm}
\hbox{\includegraphics[angle=270,width=0.5\linewidth]{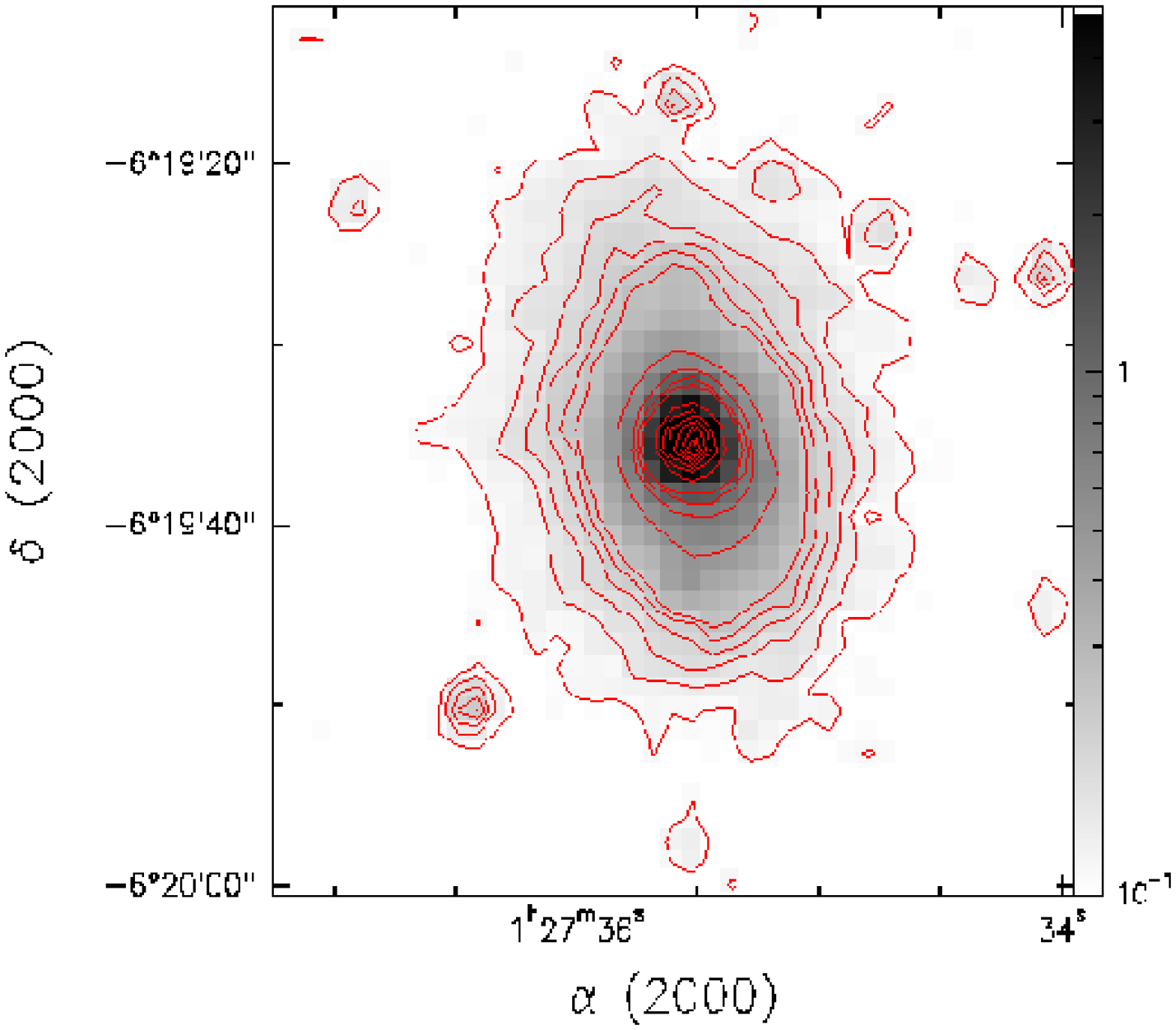}
\includegraphics[angle=270,width=0.45\linewidth]{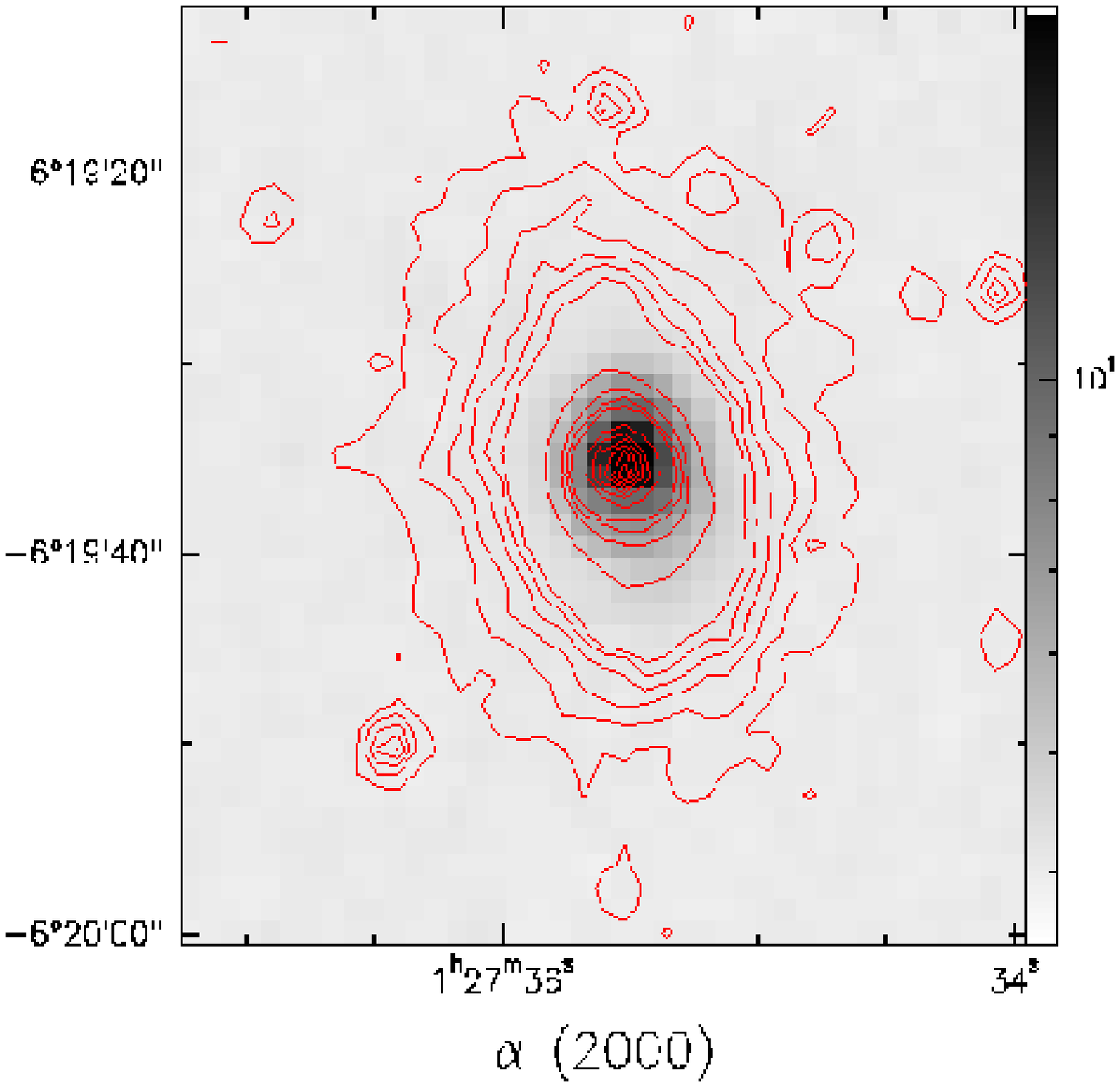} }
\caption{The 4.5\micron\ (IRAC2) image (left panel) and
the 8.0\micron\ (IRAC4) image (right panel) of \mrk.
The units given in the colorbar are MJy/sr, and the grey scale is logarithmic.
The contours in both panels correspond to the 4.5\micron\ image.
The outermost contour corresponds to 
the sky level + 10$\sigma$ ($\sigma$ = 0.0028 MJy/sr).
\label{fig:irac}}
\end{figure}

We acquired four IRAC frames, with small-scale dithering in a cycling pattern, 
giving a total of 120s spent in each of the 4.5 and 8.0 \micron\ channels.
 The individual {\it bcd} frames were processed with  
the S11.0.1 version of the SSC pipeline which 
removes the effects of
dark current, detector nonlinearity, flat field, multiplexer bleeding, 
and cosmic rays, and performs flux calibration.
The {\it bcd} frames 
were then coadded using \mopex, 
the image mosaicing and source-extraction package
provided by the SSC \citep{mopex}.
The frames were corrected for geometrical distortion and
projected onto a fiducial coordinate system
with pixel sizes of 1\farcs20, roughly equivalent to the original pixels.
Standard linear interpolation was used for the mosaics.
The final coadded 4.5 and 8.0\,\micron\ images are shown 
respectively in the left and right panels of Fig. \ref{fig:irac}.
Fig. \ref{fig:hstiracoverlay} shows 4.5\,\micron\ contours superimposed on 
the \hst/WFPC2 F791W optical image of  
\citet{thuan96} (left panel) and on the \hst\ 569W - F791W color image
from the same authors (right panel). In Fig. \ref{fig:iraccolor}, we show the 
4.5/8.0\micron\ flux ratio image on which have been overlayed the surface brightness 
level contours of the 4.5\micron\ image. To construct the flux ratio image, the 
two IRAC images have been aligned by cross-correlation after sky subtraction.

We have performed aperture photometry on the IRAC images
with the IRAF\footnote{IRAF is distributed by National Optical Astronomy 
Observatory, which is operated by the Association of Universities for 
Research in Astronomy, Inc., under cooperative agreement with the National 
Science Foundation.}
photometry package APPHOT, taking care to convert 
the MJy/sr flux units of the images to integrated flux.
The background level was determined by averaging several 
adjacent empty sky regions. We give the growth curve for both IRAC images (filled circles 
for the 4.5\micron\ image and filled squares for the 8.0\micron\ image) 
in the left panel of Fig. \ref{fig:phot}.
We have also made photometric measurements 
 of the instrument point response functions (PRFs),
provided in the \mopex\ package;
the dashed lines show the growth curve expected from a point source, normalized
to the total flux indicated by horizontal dotted lines in Fig. \ref{fig:phot}.
The 4.5 \micron\ and 8.0 \micron\ total fluxes are listed in Table 
\ref{tab:photom}. 

We have derived radial surface brightness profiles by fitting ellipses
to both IRAC images.
The ellipse centers were held fixed, but the ellipse shape and orientation was
allowed to vary with radius.
The surface brightnesses were put on the Vega magnitude scale 
by using the photometric calibration
of \citet{reach05}.
Figure \ref{fig:profile} (left panel) shows the radial profiles in units of
[4.5] and [8.0] magnitudes.
The average disk color of 1.5 corresponds to a flux ratio 4.5/8.0 of 0.7.
The red circumnuclear maximum in the lower panel at a radius of 
$\sim$ 4\arcsec\ is an artifact due to the different
IRAC point-response functions at the two wavelengths.
This effect can also be seen in Fig. \ref{fig:iraccolor} where the crowding
of the 4.5\,\micron\ contours from the point source
coincides with a relatively low 4.5/8 flux ratio.

\subsection{\label{sec:mipsdata}MIPS Imaging}

We have obtained for \mrk\ 
a total of 24 frames at 24\micron, 56 frames at 70\micron, 
and 104 frames at 160\micron, using ramp times of 3s, 10s, and 10s 
respectively, 
with 1, 2, and 6 cycles in the three channels.
The individual {\it bcd} frames were processed by the S11.4.0 version of the SSC pipeline,
which includes  
dark current subtraction, flat-fielding, and flux calibration.
As for the IRAC images, we processed the dithered {\it bcd} frames in 
the spatial domain with \mopex.
Geometrical distortion was corrected before projecting the frames onto 
a fiducial coordinate system with pixel sizes of 1\farcs2 for MIPS-24,
roughly half of the original pixel size of 2\farcs5.
Pixel sizes of the final mosaics at 70 and 160\micron\ are also 
approximately half
of the originals, i.e. 4\farcs95 at 70\micron\ and 8\farcs0 at 160\micron.
Unlike the IRAC coadds,
we incorporated the sigma-weighting algorithm because we found
it gave less noisy MIPS mosaics than without.
Standard linear interpolation was used in all cases.
In all three channels, our  post-pipeline \mopex\ mosaics are superior to 
those provided by the automated post-pipeline reduction.
The final coadded images are shown in Fig. \ref{fig:mipsoverlay} as 
contours superimposed on
the IRAC 4.5 \micron\ image. We have performed aperture photometry on the 
MIPS images in the same way as on the IRAC images. The growth curves for the three 
bands (filled circles for the 24 \micron\ band, open squares for the  70 \micron\ band
and crosses for 160 \micron\ band) are given in the right panel of 
Fig. \ref{fig:phot}. The dashed lines represent the point-response function in each band.
The 24 \micron, 70 \micron\ and 160 \micron\ total fluxes are listed in Table 
 \ref{tab:photom}. 

\subsection{\label{sec:irsdata} IRS Spectra}

Spectroscopy was performed in the staring mode with the Short Low module 
in both orders (SL1, SL2)
and with both Short and Long High-resolution modules (SH, LH). 
The data consist of  
low-resolution spectra with a wavelength range from 
5.2 to 14.5\,\micron\ and a spectral resolution of $\simeq$64-128, and 
of high-resolution spectra with a wavelength range 
from 9.6 to 37.2\,\micron\ and a spectral resolution of $\simeq$600.
Total integration times of 30s$\times$4 cycles were 
obtained for SH, 14s$\times$8 for LH,
6s$\times$6 for SL1, and 6s$\times$6 for SL2.
The central region of 
Mrk 996 was centered in the slits by peaking up on a nearby 2MASS star.
Individual {\it bcd} frames have been processed by the S13.2.0 version of the 
SSC pipeline, which provides ramp fitting, dark current 
subtraction, droop and linearity corrections,
flat-fielding, and wavelength and flux calibrations\footnote{See the IRS
Data Handbook, \url{http://ssc.spitzer.caltech.edu/irs/dh}.}.

However, the pipeline does not include background subtraction.
Hence, for the low-resolution spectra, we constructed a coadded 
background frame from the 
{\it bcd} observations 
with the source in the opposing nod and off-order positions
\citep[see also][]{weedman05}. 
For the high-resolution SH and LH spectra, 
a 2D background image could not be 
constructed because of the small size of the slit.
Therefore we subtracted the background from the SH and LH observations
using the one-dimensional (1D) spectra.
The level of the background was determined by minimizing the difference
between the SL and SH$+$LH spectra over their overlap region ($\la$1\,\micron),
and the shape of the background was given by Spot.

We extracted the source spectra with \spice, the post-pipeline IRS package
provided by the SSC.
To maximize the calibration accuracy,
the automatic point-source extraction window was used for all modules.
For the SL spectra, this gives a 4-pixel (7\farcs2) length at 6\,\micron, and
an 8-pixel one (14\farcs4) at 12\,\micron; 
the slit width is 3\farcs6 for both SL modules.
At high resolution, the \spice\ extraction is performed over the entire slit
(4\farcs7$\times$11\farcs3 SH; 11\farcs1$\times$22\farcs3 LH).
The individual spectra were then box-car smoothed to a resolution element, 
and clipped in order to
eliminate any remaining spikes in the high-resolution data.
The spectra were also examined for bad pixel removal.  
Finally, the two spectra for each module (one for each nod position) 
were averaged.
The  final averaged IRS 
spectra obtained in 
the short-wavelength low-resolution mode (SL) and in both short and 
long wavelength  
high-resolution modes (SH, LH) are shown in Fig. \ref{fig:irs}. A
blow-up of the spectral region around the [O {\sc iv}] line is shown in 
Fig. \ref{fig:oiv}.

\section{Complementary Observations}

\subsection{UKIRT Imaging}

We have also acquired $J$ (1.2 \micron), $H$ (1.6 \micron), and $K$ (2.2 \micron) 
images of \mrk, 
with the 3.8 m UKIRT1 equipped with IRCAM3 
as part of our ongoing project of NIR imaging and spectroscopy of BCDs. 
The IRCAM3 plate scale is 0\farcs28 pixel$^{-1}$, 
with a total field of view of 72\farcs0 $\times$ 72\farcs0. Source and empty 
sky positions were alternated, beginning and ending each observing sequence 
with a sky position. Before the beginning of each sequence, dark exposures 
were acquired with the same parameters as the subsequent science frames. Total 
on-source integration times were 1040s in $K$, 
480s in $H$, and 240s in $J$. Individual frames were first dark-subtracted 
and then flat-fielded with the average of adjacent empty sky frames, after editing 
them for stars (to avoid``holes'' in the reduced frames) and applying an 
average $\sigma$ clipping algorithm. 
The reduced frames were then aligned and averaged. All data reduction was 
carried out in the IRAF environment.

     Photometric calibration was performed by observing standard stars from 
the UKIRT 
Faint Standard List \citep{hawarden01} before and after the source 
observations. Each standard star was measured in several different positions 
on the array and flat-fielded by dividing the clipped mean of the remaining 
frames in the sequence. To correct the standard-star photometry for 
atmospheric extinction, we used the UKIRT mean extinction coefficients of 
0.102, 0.059, and 0.088 mag per air mass for $J$, $H$, and $K$, respectively. 
Formal photometric accuracy, as measured by the dispersion in the 
standard-star magnitudes, is 0.025 mag in $J$ and $H$ and 0.04 mag in $K$.
Color images were derived by registering the images to a few tenths of a pixel 
with a cross-correlation algorithm, then subtracting the magnitude images. 

As for IRAC, we have fitted ellipses to the $J$, $H$ and $K$ images of \mrk, 
and $J$, $H$, and $K$ elliptically-averaged radial profiles have been derived.
The $K$ surface brightness profile, which reaches 
a limiting surface brightness of $\sim$ 24 mag arcsec$^{-2}$,
is shown in the top right panel of  Fig. \ref{fig:profile}.
The lower three panels show the \jh, \hk\ and \jk\ radial color profiles,
constructed from the $J$, $H$ and $K$ surface brightness profiles. In the last 
three panels,
the dotted line indicates the average color of the underlying disk. 

\subsection{\label{sec:mmt} MMT Optical Observations}


As the $HST$ optical spectra \citep{thuan96} 
have somewhat low spectral resolution 
($\sim$ 6\AA\ in the blue or $\sim$ 350 km s$^{-1}$), higher 
resolution optical spectra are needed to allow for a deconvolution 
of the line profiles into narrow and broad components and better 
study the gas motions in the inner and outer regions.
A new high signal-to-noise ratio 
optical spectrum of Mrk 996 was obtained with the 6.5 m
MMT on the night of 2006 December 15.  
Observations were made with the Blue Channel of the MMT spectrograph. We used
a 1\farcs5$\times$180\arcsec\ slit and a 800 grooves/mm grating in first order.
The above instrumental set-up gave a spatial scale along the slit of 0\farcs6
pixel$^{-1}$, a scale perpendicular to the slit of 0.75\AA\ pixel$^{-1}$,
a spectral range 3200 -- 5000\AA\ and a spectral resolution of 3\AA\ (FWHM).
The slit was oriented along the parallactic angle and the total exposure 
time was 25 minutes. The observations were broken up into 3 subexposures to 
avoid saturation of the brightest lines. 
The seeing was about 0\farcs8. 
The Kitt Peak IRS spectroscopic standard stars Feige 110 and G191B2B 
were observed for flux
calibration. Spectra of He-Ar comparison arcs were obtained 
after the observations to calibrate the wavelength scale. 

The data reduction procedures are the same as described in 
\citet{thuanizotov05}.
The two-dimensional spectra were bias subtracted and 
flat-field corrected using IRAF.
We then use the IRAF
software routines IDENTIFY, REIDENTIFY, FITCOORD, TRANSFORM to 
perform wavelength
calibration and correct for distortion and tilt for each frame. 
 Night sky subtraction was performed using the routine BACKGROUND. 
The level of
night sky emission was determined from the closest regions to the galaxy 
that are free of galaxian stellar and nebular line emission,
 as well as of emission from foreground and background sources.
A one-dimensional spectrum was then extracted from the two-dimensional 
frame using the APALL routine. Before extraction, the three 
distinct two-dimensional spectra of Mrk 996 
were carefully aligned using the spatial locations of the brightest part in
each spectrum, so that spectra were extracted at the same positions in all
subexposures. 
We then summed the individual spectra 
from each subexposure after manual removal of the cosmic rays hits. 
The spectra obtained from each subexposure
were also checked for cosmic rays hits at the location of strong 
emission lines, but none were found.

Particular attention was paid to the derivation of the sensitivity curve. 
It was obtained by 
fitting with a high-order polynomial the observed spectral energy 
distribution of the two standard stars. 
Because the spectra of these stars have only a small number of a 
relatively weak absorption features, their spectral energy distributions are  
known with good accuracy ($\la$ 1\%). 
Moreover, the response function of the CCD detector is smooth, so we could
derive a sensitivity curve with an accuracy better than 1\% over the
whole blue optical range.

The resulting spectrum of Mrk 996 is shown in Fig. \ref{fig:sp}. 
The spectrum
is very rich in emission lines, which we have identified and labeled.
Because of its higher spectral resolution and signal-to-noise ratio, 
the MMT spectrum shows many more weak emission lines than the 
{\sl HST} spectrum, confirming, strenthening and extending 
many of the findings of \citet{thuan96}: 1) the He {\sc i} 
line intensities such as that of the He {\sc i} 
$\lambda$4471 emission line are unusually large.
There are many weak He {\sc i} lines that are not ordinarily seen 
such as the $\lambda$3820, $\lambda$4121, $\lambda$4143 and $\lambda$4388
lines; 
2) the widths of the emission lines of high-ionization stages 
are broader than those of low-ionization stages; 3) there are strong,
broad Wolf-Rayet bumps at $\lambda$4640 and $\lambda$4686,
suggesting the presence of a large WN stellar population. 
\citet{thuan96} have also 
detected a C {\sc iv} WR bump at $\lambda$5808, indicative of WC stars.
In addition, the spectrum shows the presence of numerous weak [Fe {\sc ii}] 
and [Fe {\sc iii}] lines. The spectrum goes far into the blue, which  
allows us to check for the presence of the 
 [Ne {\sc v}] $\lambda$3426 line, an indicator of hard ionizing radiation 
since its ionization potential is 7.1 Rydberg. It is not seen. 
The [Fe {\sc v}] $\lambda$4227 
with an ionization potential of 4 Rydberg is also absent.
The He {\sc ii} $\lambda$4686, also with an ionization potential of 4 Rydberg,
is seen but its broad width suggests that it originates in stellar 
winds of WR stars, not from the ionized interstellar gas. 
The Balmer jump, clearly seen at 
$\lambda$3646, is useful for determining the temperature of the H$^+$ 
zone \citep{guseva06}. 

   The observed line fluxes $F$($\lambda$) normalized to $F$(\hb) and multiplied by 100 
and their errors, are given in 
Table \ref{tabint}. They were measured using the IRAF SPLOT routine. The 
line flux errors listed include statistical errors derived with SPLOT from 
non flux-calibrated spectra, in addition to errors introduced in the standard 
star absolute flux calibration. Since the differences between the 
response curves derived for the two standard stars are not greater than 1\%, 
we set the errors in flux calibration to 1\% of the line fluxes. 
 The line fluxes were corrected for both reddening \citep{w58}
 and underlying hydrogen stellar absorption derived simultaneously by an 
iterative procedure as described in \citet{itl94}. The corrected line fluxes 100 $\times$
$I$($\lambda$)/$I$(\hb) and equivalent widths EW($\lambda$) are also given in 
Table \ref{tabint}. 

Comparison of the relative line intensities given in 
Table \ref{tabint} with those measured by \citet{thuan96} shows general 
good agreement. The largest deviation concerns the [O {\sc ii}] $\lambda$3727 
line which is about 1.5 times larger. We attribute this difference 
to an aperture effect: the {\sl HST} aperture is circular with a diameter 
of 0\farcs86 while the MMT slit width is 1\farcs5. Since the low-ionization 
ions are produced in the outer less dense regions, a larger aperture would 
have a larger [O {\sc ii}] to \hb\ ratio.

\section{Imaging Results: The Infrared Morphology of \mrk}

We now use the UKIRT near-infrared and the 
Spitzer IRAC and MIPS mid-infrared imaging data
to study the origin of the infrared emission in \mrk. 

\subsection{An exponential disk of $\ga$ 1 Gyr old stars} 
 
The top left and right panels of Fig. \ref{fig:profile} show respectively 
the IRAC and $K$-band surface 
brightness profiles of \mrk, which are very similar to the $V$ and $I$ profiles given by 
\citet{thuan96}. In the radius range $r$ $\ge$ 4\arcsec, they can be fit by 
an exponential law of the form $\mu_K$ = 17.1 + 0.29 $r$(\arcsec), 
$\mu_{4.5}$ = 16.7 + 0.25 $r$(\arcsec), 
corresponding to scale lengths $\alpha$$_K^{-1}$ $\sim$ $\alpha$$_{4.5}^{-1}$ 
$\sim$0.4\,kpc, similar 
to the $V$ and $I$ scale lengths. This means that for $r$ $\ge$ 4\arcsec, there 
is no $V-K$ color gradient, and hence no stellar population change in 
the underlying extended low surface brightness component. This 
is also consistent with the constant $V-I$ (= 0.80) \citep{thuan96}, 
\jh\ (=0.57) and \hk\ (=0.14) colors observed  
for $r$ $\ge$ 4\arcsec\ (panels 2 and 3 of Fig.\ref{fig:profile}). 
These colors have been corrected for Galactic extinction $A_V$=0.146\,mag
\citep{schlegel98} with the \citet{cardelli89} interstellar extinction curve.

The colors correspond to a coeval stellar population formed about 1\,Gyr ago 
according to the models of \citet{vazquez05}.
This age is fairly well constrained by the models for the optical colors,
but in the NIR, models are still unable to completely predict broadband
colors for intermediate-age stellar populations at sub-solar metallicities
\citep[e.g.,][]{origlia99,mouhcine02,vazquez05}.
The influence of red supergiants and asymptotic giant branch stars at
ages of $\la$1\,Gyr increase with decreasing metallicities, and predicted
NIR colors tend to be too blue in \jh\ and too red in \hk\ as compared to
the observed ones.
In any case, the observed NIR colors are clearly not those of 
old stellar populations in spiral disks, dominated by low-mass giants
\citep[e.g.,][]{dejong96,peletier97,hunt97}.
The oldest stars are probably older than 1\,Gyr as 
the star formation in \mrk\ is likely more spread out in time
as compared to the instantaneous star formation 
models we have used. 

\subsection{A very red nucleus}   
   
Just like the $V$ and $I$ surface brightness profiles \citep{thuan96}, 
the $K$ profile shows a very steep gradient toward the nucleus.
While the $V-I$ profile becomes rather blue, in the near-infrared 
the striking feature of the nuclear region is its  
extremely red \jk\ color, equal 
to $\simgt$1.8, as compared to the \jk\ color of the disk equal only to $\sim$0.7. 
Examination of the lower right panels of
Fig. \ref{fig:profile} shows that the redness of the 
nuclear region comes mainly from its \hk\ color (equal to $\sim$1.0 as compared  
to the \hk\ = 0.1 of the disk), 
and less from its \jh\ color. 
The red colors cannot be produced by gaseous emission as 
pure gaseous colors are \jh\ = 0.0 and \hk\ = 0.6 \citep{hunt03}.    
The redness is more important at longer wavelengths, 
suggesting that it may be caused by a very hot dust component 
in the nucleus of \mrk,
similar to, although more extreme than, other star-forming galaxies
\citep{hunt02}. 

Since the brightness gradient toward the nucleus is very steep, we have
checked whether variable seeing for different exposures  
can be responsible for the very red colors observed. We found that 
the red \jk\ color of $\geq$2 shown in the color profile is  
slightly enhanced because of the better
seeing in the $K$ band image than in the $J$ band image.
However, the very red \hk\ color cannot be attributed to seeing effects.
The seeing was 0\farcs9 (FWHM) during the $H$ band observations, and
0\farcs8 in the $K$ band observations.
We have convolved the $K$-band image to the $H$-band seeing, and 
produced a new \hk\ color image. The
nuclear color in this new image is $\simeq$1.0, 
just as in the profile shown in Fig. \ref{fig:profile}.
We have also checked the \jh\ color. The 
registration process necessary to align the images 
already degrades the $H$ band image to the seeing of the $J$-band image 
of 1\farcs0.
The resulting nuclear \jh\ color is $\sim$0.8.
Thus, 
the nuclear colors are still very red after correction for seeing effects:
\hk\ = 1.0, and \jh\ = 0.8.

Fig. \ref{fig:jhk} shows in a (\jh, \hk) diagram, as filled rectangles connected by a 
dotted line and from right to left,  
the colors of the nucleus, of a circumnuclear location, 
and of a disk location in \mrk,
together with 
those of IR-bright starburst galaxies \citep{hunt02} (open stars), 
of selected BCDs (asterisks) including NGC 5253, SBS 0335--052 \citep{hunt02},
and Haro 3 \citep{johnson04}, and of Seyfert galaxies (filled circles)
\citep{alonso01}.
We have also shown in Fig. \ref{fig:jhk}
mixing curves that illustrate how adding various physical 
components to the stellar component modify the $JHK$ colors of the disk
of \mrk, corrected for the K effect and Galactic
extinction. The five solid lines from top to bottom show respectively 
the effects of dust extinction, of a hot dust component at 600\,K and 1000\,K, 
of free-free gas emission and of an A-star
population. 
It can be seen that the $JHK$ colors of the disk of \mrk\ are similar 
to the colors of other BCDs, and
slightly bluer than those of normal (metal-rich) starbursts, but that 
the \hk\ color of the nucleus of \mrk\ is considerably redder. 
While other BCDs show a clear contribution from ionized gas,
the $JHK$ colors of \mrk\ show a significant component of 
hot dust, with a dust temperature between 600 and 1000\,K.


What is the appearance of the nucleus at longer wavelengths? 
Examination of the MIPS images (Fig. \ref{fig:mipsoverlay}) shows that 
it is clearly detected at 24, 70 and 160\,\micron.
The emission at 24 and 70\,\micron\ is virtually point-like, as shown by  
the characteristic Airy rings seen in the top two panels of 
Fig. \ref{fig:mipsoverlay}, and from the growth curves in Fig. \ref{fig:phot}.
However, the emission at 160\,\micron\ appears to be more extended, as is 
also evident from its growth curve, suggesting a significant contribution from 
the galaxy disk.     
   
\subsection{The circumnuclear PAH emission} 
 
\begin{figure}
\vspace{0.2cm}
\centerline{ \includegraphics[angle=270,width=0.5\linewidth]{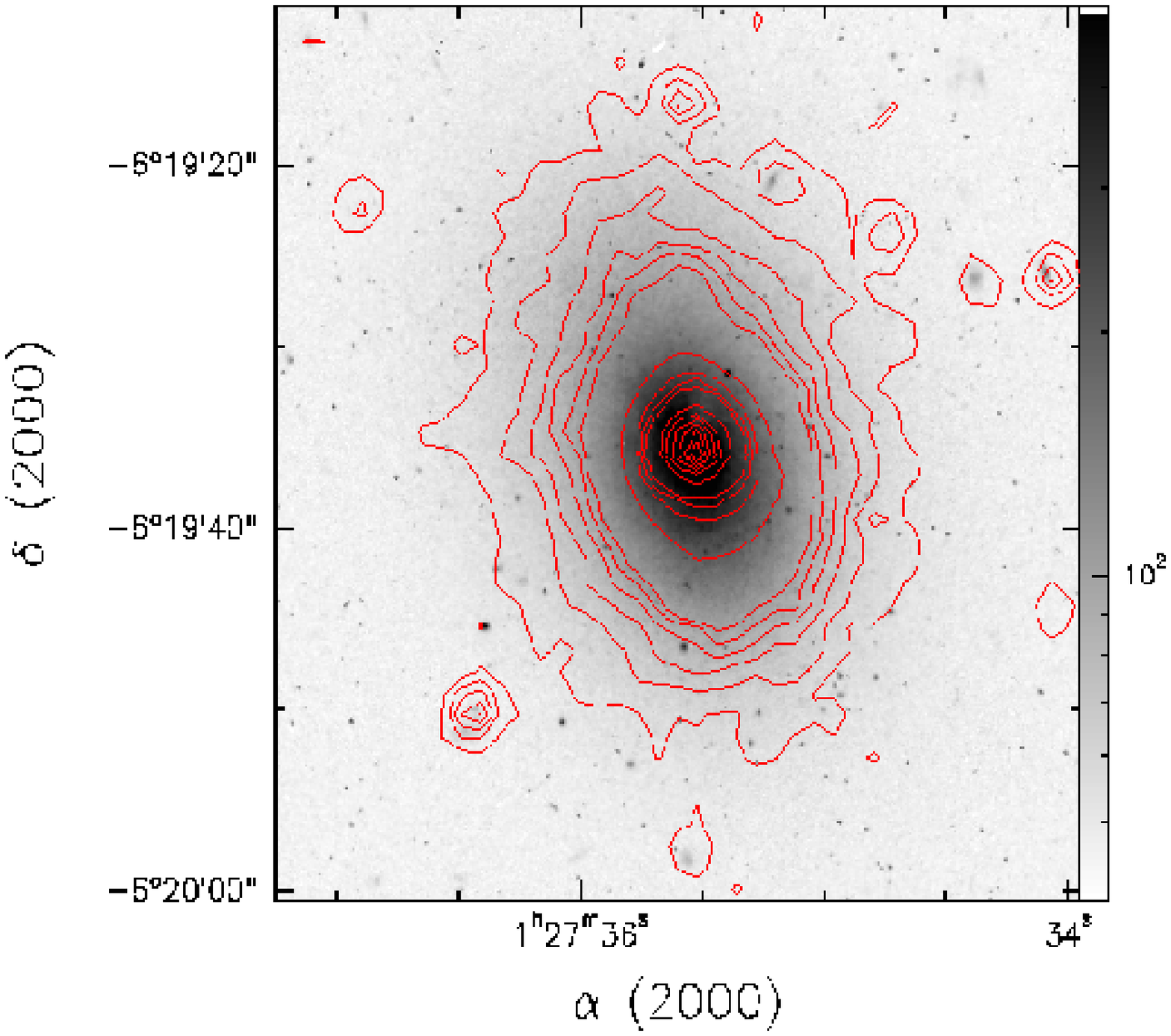} 
\includegraphics[angle=270,width=0.45\linewidth]{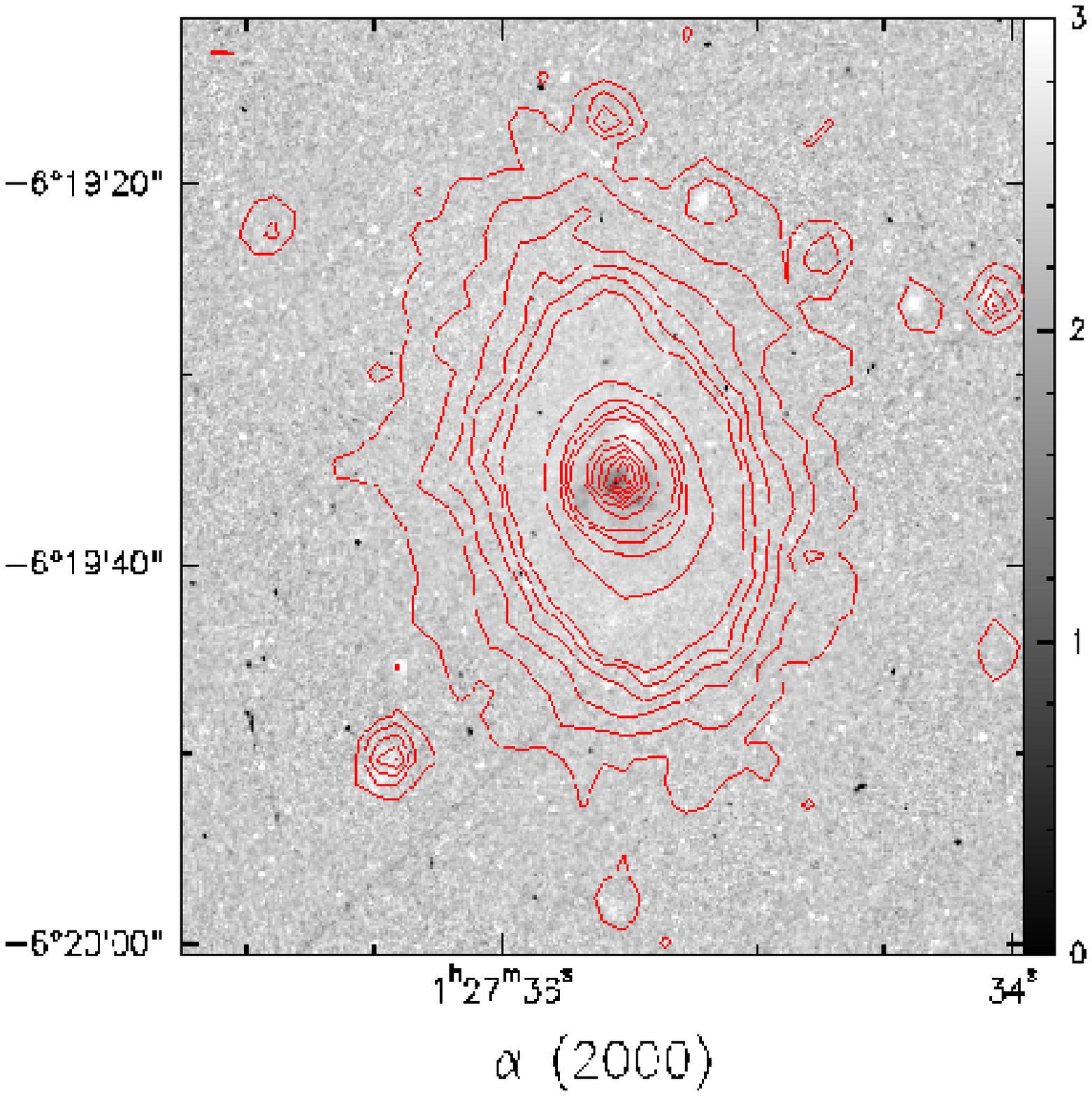} }
\caption{The 4.5\micron\ IRAC image of \mrk\  
contoured on: (left panel) the \hst /WFPC2 F791W image \citep{thuan96}
and (right panel) the \hst\ F569W$-$F791W color image \citep{thuan96}.
The units in the color bar are counts (left) and magnitudes (right), 
and the grey scale is logarithmic.
The contours in both panels correspond to the 4.5\micron\ image (see Fig. \ref{fig:irac}). The outermost contour corresponds to  
the sky level + 10$\sigma$ ($\sigma$ = 0.0028 MJy/sr).
The IRAC 4.5\micron\ brightness peak is centered on the 
blue nucleus (F569W$-$F791W$\simlt$0) of the galaxy, rather than on 
the red dust patch (F569W$-$F791W$\sim$3) to the northwest.
\label{fig:hstiracoverlay}}
\end{figure}

\begin{figure}
\vspace{0.2cm}
\centerline{ 
\includegraphics[angle=270,width=0.9\linewidth]{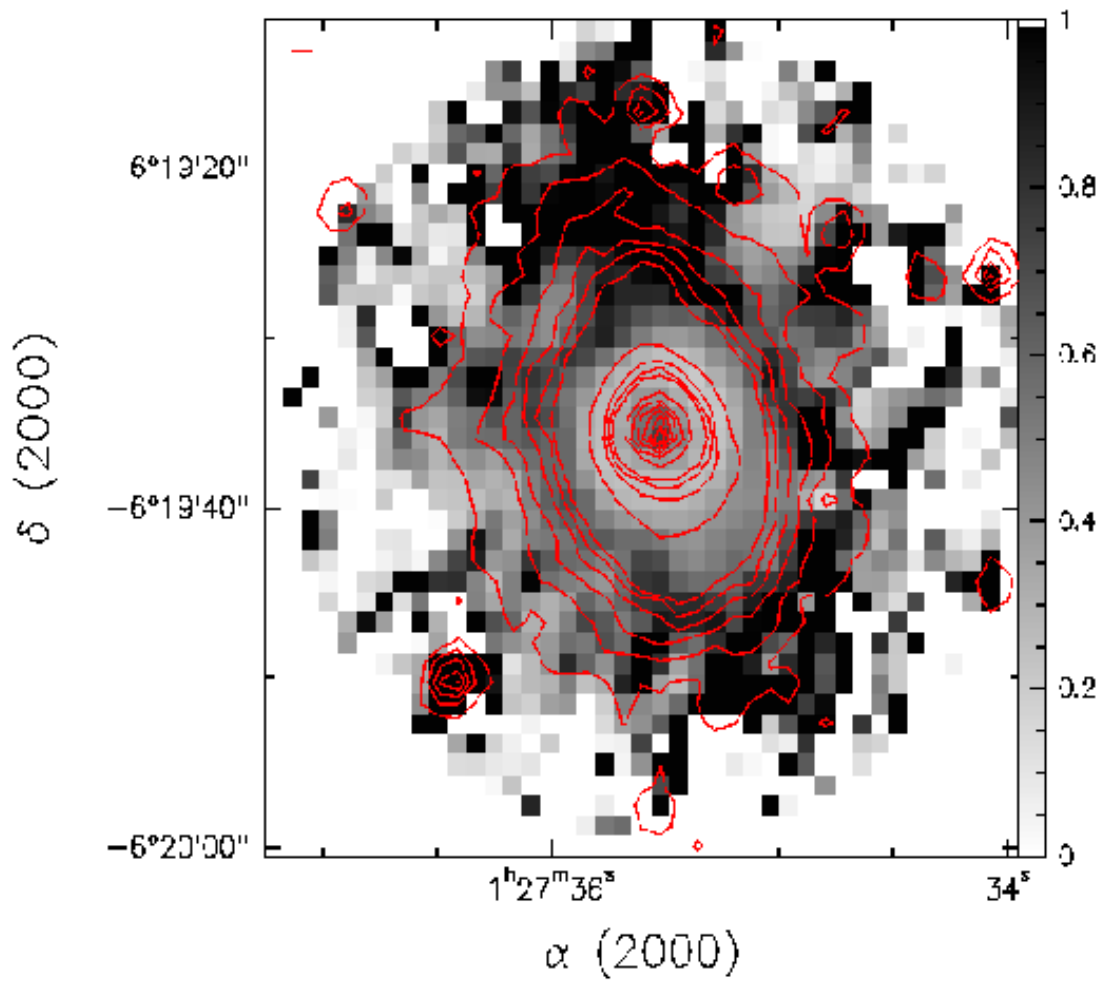} }
\caption{The 4.5\micron\ IRAC image of \mrk\ 
contoured on the IRAC 4.5/8.0\micron\ flux ratio image.
The flux ratio units given in the color bar are in MJy/sr.
As in Fig. \ref{fig:irac}, the outermost contour corresponds to  
the sky level + 10$\sigma$ ($\sigma$ = 0.0028 MJy/sr).
The local maximum of the 4.5/8.0\micron\ flux ratio
is on top of the 4.5\micron\ brightness peak, which also coincides
with the blue nucleus (F569W$-$F791W$\simlt$0) of the galaxy.
There is very little 8\,\micron\ emission outside the circumnuclear
region, so the color ratio map is considerably more
noisy in the outer parts of the disk.
\label{fig:iraccolor}}
\end{figure}

\begin{figure}
\vspace{0.2cm}
\hbox{\includegraphics[angle=0,width=0.5\linewidth]{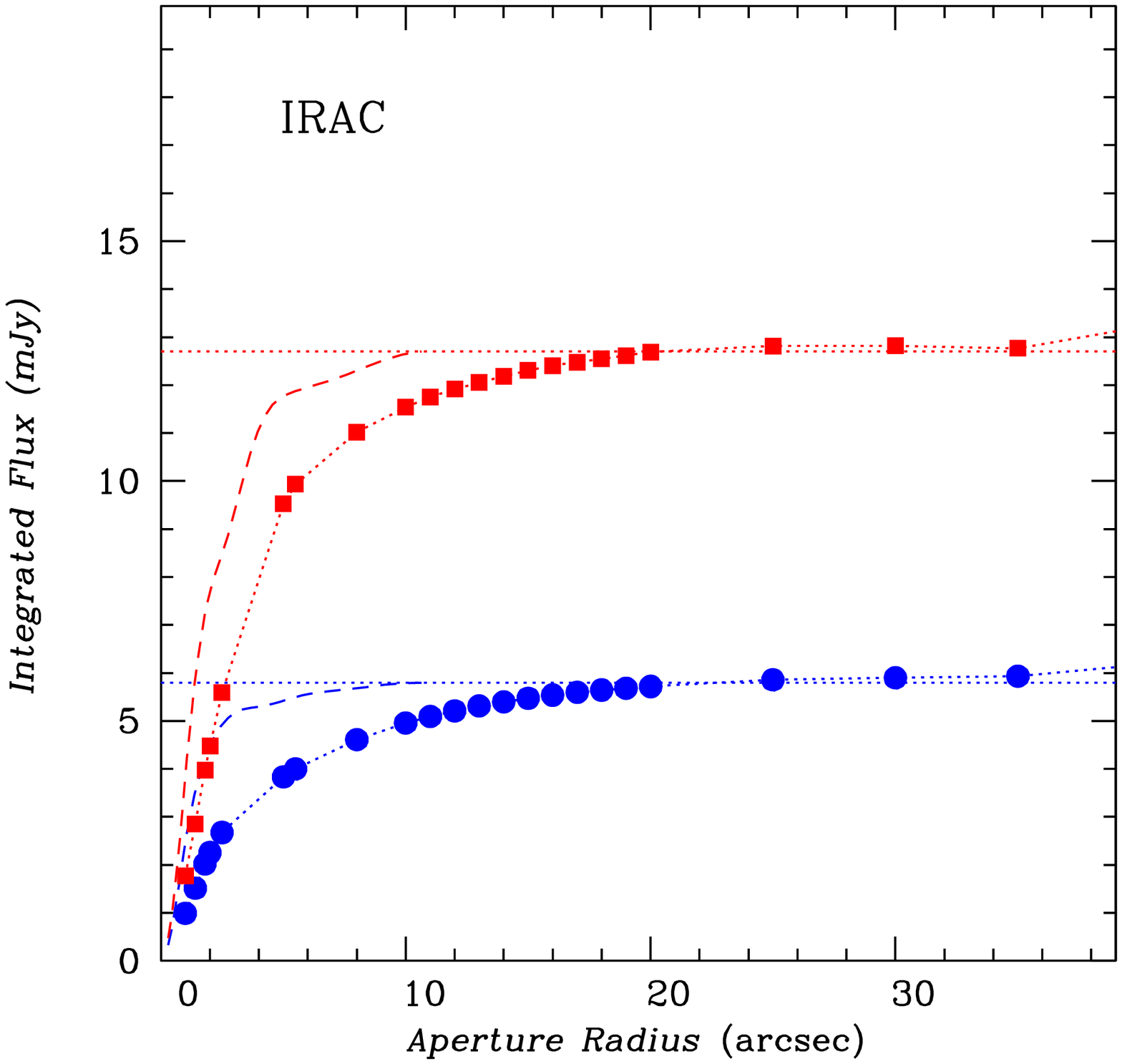}
\includegraphics[angle=0,width=0.5\linewidth]{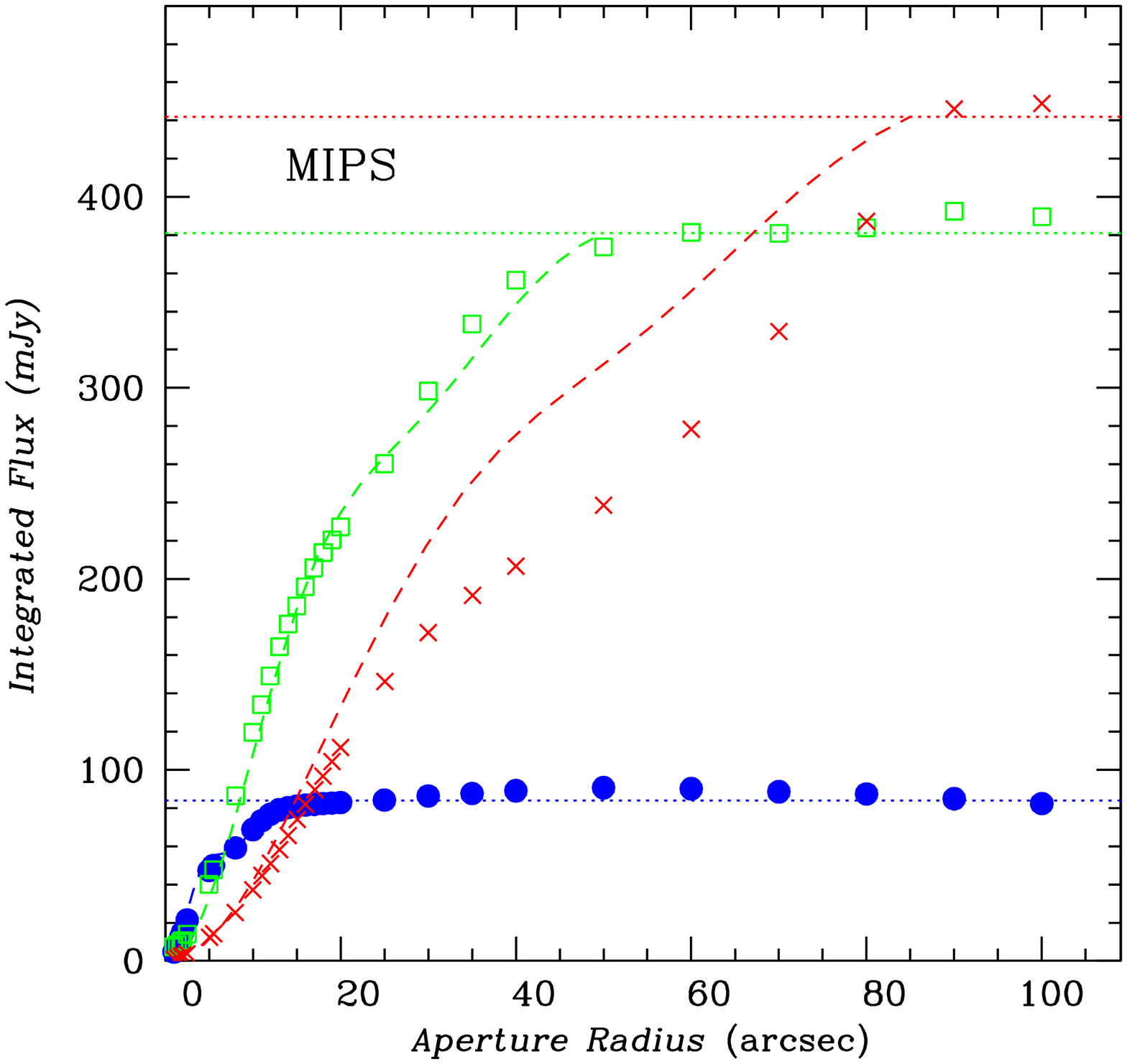} }
\caption{Growth curves for IRAC (left panel) and MIPS (right) photometry of \mrk. 
In both panels, for each wavelength, the horizontal dotted line shows the adopted total
flux of \mrk, and
the dashed line represents the point-source response function.
In the left panel, the 4.5 \micron\ IRAC photometry is indicated by filled circles,
and the 8.0 \micron\ photometry by filled squares.
In the right panel, the MIPS-24\micron\ photometry is shown by filled circles,
the MIPS-70\micron\ photometry by open squares, and the MIPS-160\micron\ photometry
 by $\times$. 
\mrk\ is extended at shorter wavelengths and possibly at 160\micron, but
appears essentially point-like at 24 and 70\micron. 
\label{fig:phot}}
\end{figure}

\begin{figure}
\vspace{0.2cm}
\hbox{
\includegraphics[angle=0,width=0.5\linewidth,bb=48 285 590 679]{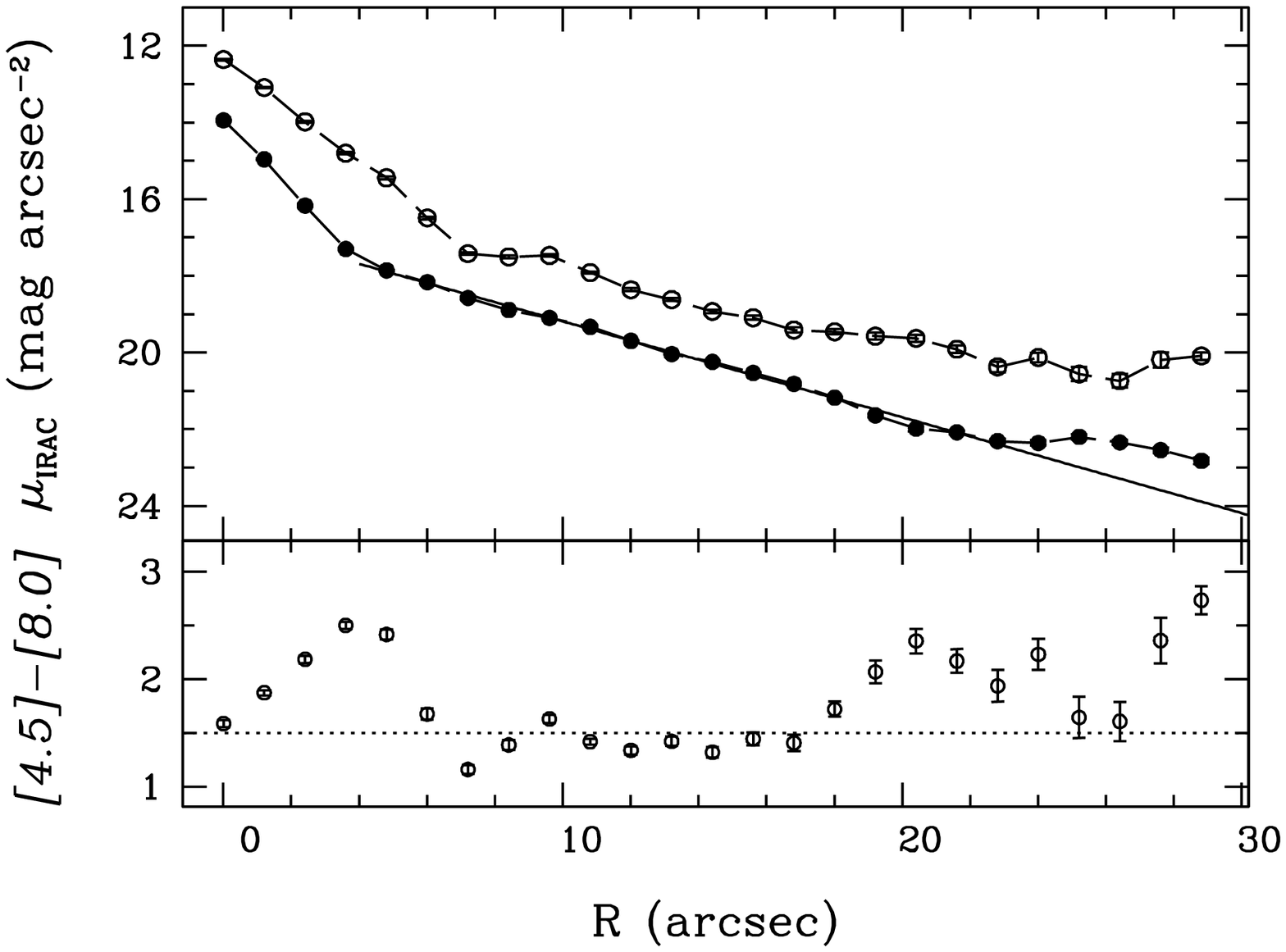} 
\includegraphics[angle=0,width=0.5\linewidth,bb=48 140 566 653]{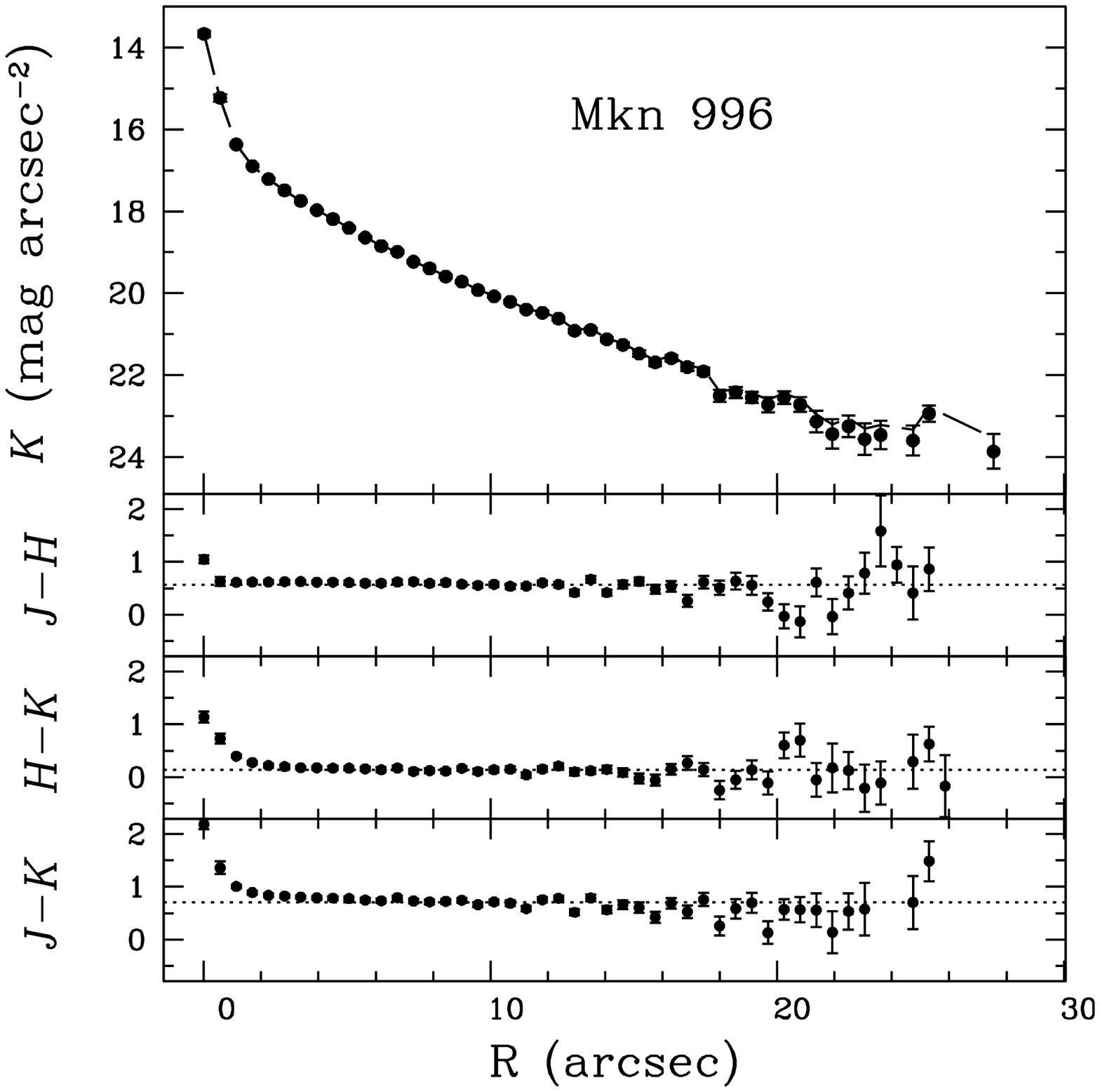} }
\caption{(a) Left panels: (top) IRAC 4.5 \mic\ and 8.0 \mic\ 
surface brightness profiles,
averaged over elliptical contours;(bottom)
color profile in units of [4.5]-[8.0] magnitudes.
In the top panel, the 4.5 \mic\ 
profile is shown as filled circles, and the 8.0 \mic\
one as open circles.
The exponential disk fit for $R>4$\,\arcsec\ described in the text
is shown as a solid line in the top panel,
and the average disk color as a dotted line in the bottom one.
The circumnuclear red ``peak'' at $R$ $\sim$ 4 \arcsec\ 
is an artifact due to the mismatch of the IRAC 4.5 \mic\ and 8.0 \mic\ 
point-response functions.
(b) Right panels: The {\it K}-band surface brightness profile of \mrk, averaged over 
elliptical contours (top panel). The three bottom panels show the \jh, \hk, and \jk\
color profiles.
In these panels, the horizontal dotted line indicates the average
color of the disk.
The \jk\ color of the nuclear region is extremely red; most of this redness 
comes from the \hk\ color and less from the \jh\ color. 
\label{fig:profile}}
\end{figure}

\begin{figure}
\vspace{0.2cm}
\hbox{\includegraphics[angle=270,width=0.5\linewidth]{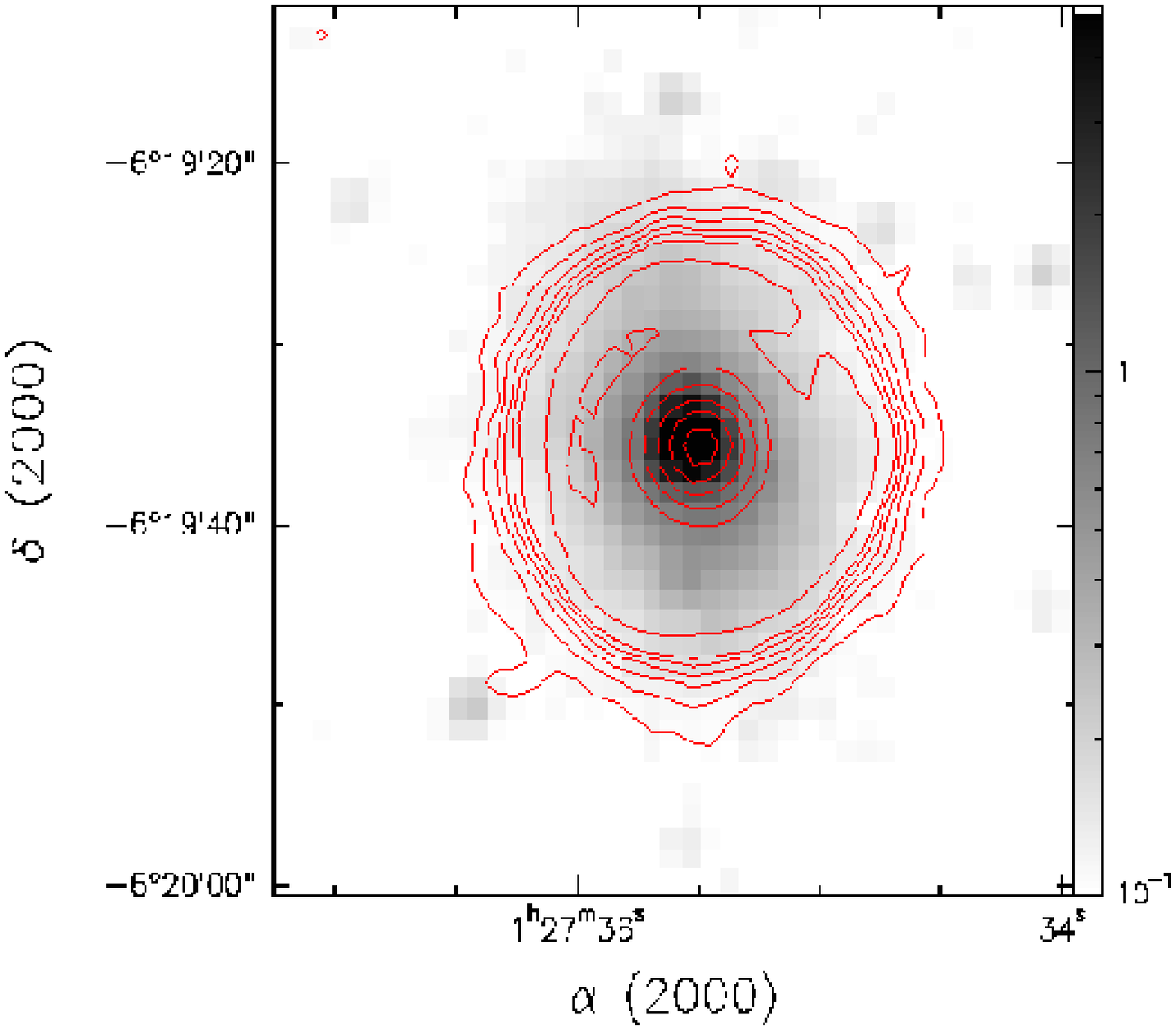}
\includegraphics[angle=270,width=0.5\linewidth]{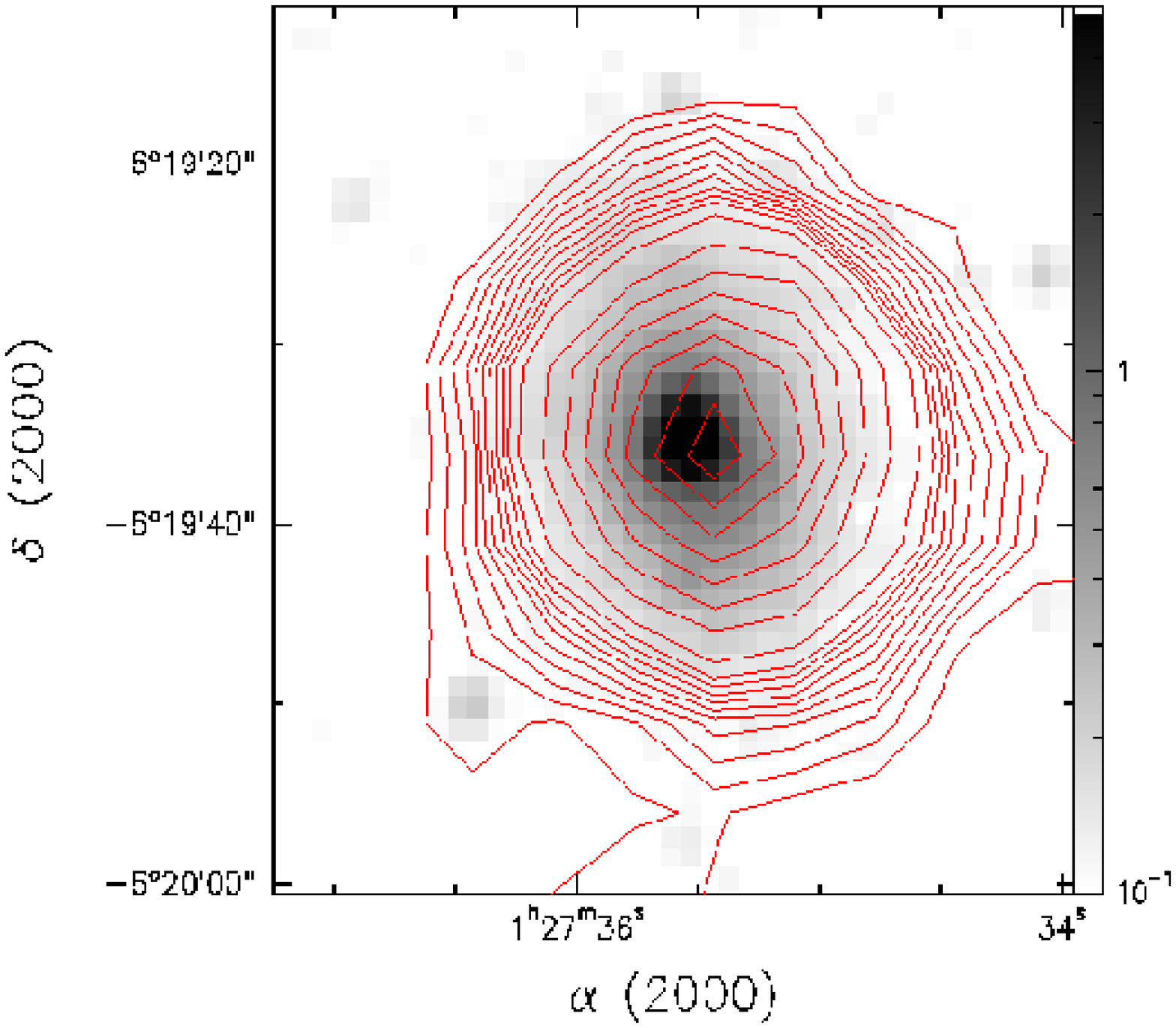} }
\hbox{\includegraphics[angle=270,width=0.5\linewidth]{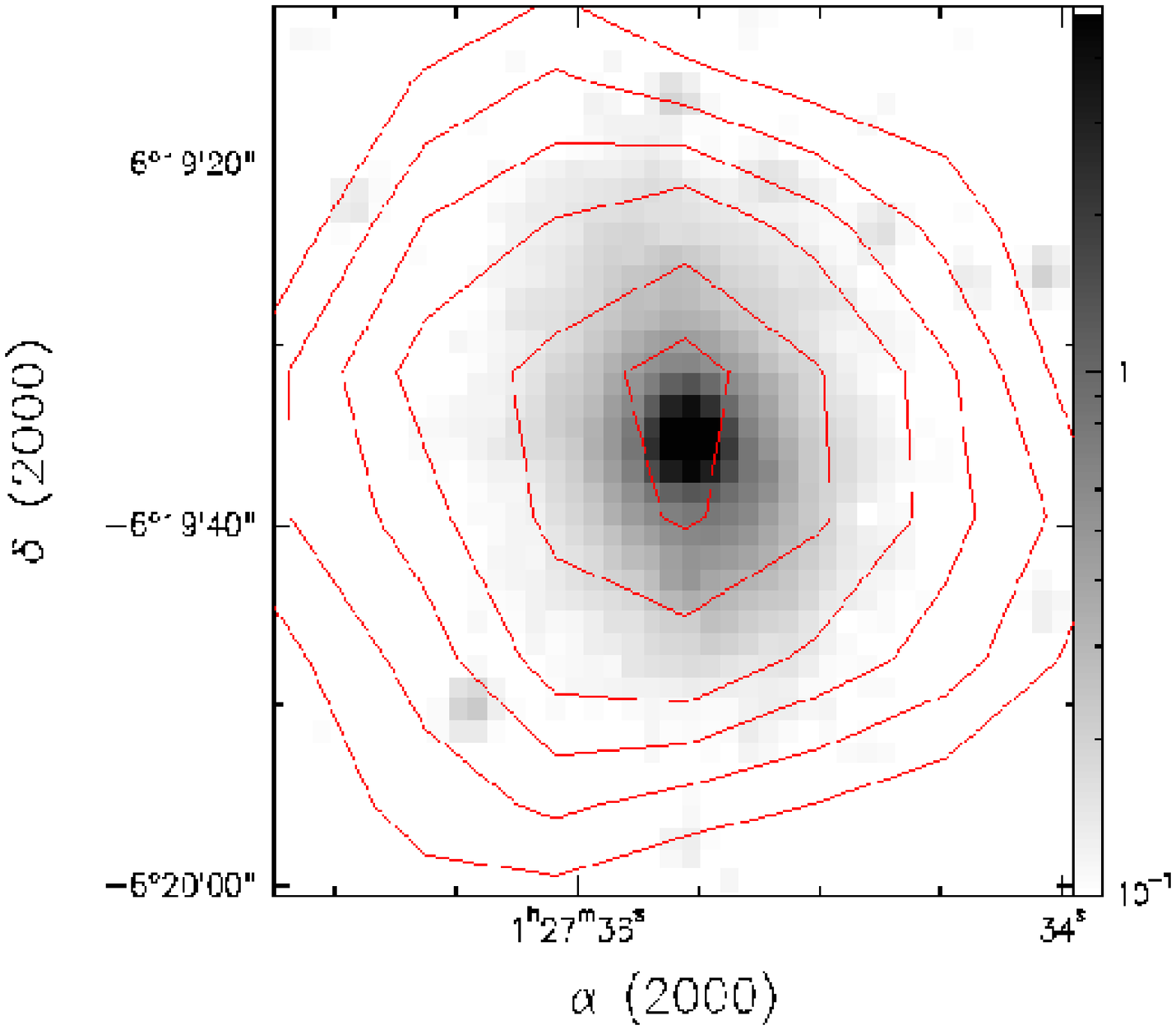} }
\caption{The 24 \micron\ (top left panel), 70 \micron\ (top right), and 160 \micron\
(bottom) MIPS images of \mrk\ 
contoured on the 4.5\micron\ IRAC image.
Contours run from the sky level + $n$$\sigma$ to the maximum intensity,
where $n$\,=\,12, 5, and 3 for MIPS-24, MIPS-70, and MIPS-160, respectively.
The signature of the Airy ring, associated with a dominant point source, is evident
in the 24 and 70\,\micron\ contours. 
\label{fig:mipsoverlay}}
\end{figure}

Examination of Fig. \ref{fig:irac} and of the growth curves shows that both the 
4.5 and 8.0 \micron\ emissions are more extended than that of a point source. 
While the 8\,\micron\ emission comes mainly 
from the nuclear and inner-disk regions (out to 
a radius of $\sim$10\arcsec\ or 1.1\,kpc), the 4.5\,\micron\ image shows 
extended emission out to a radius of $\sim$ 20\arcsec\ or 2.2\,kpc, 
comparable in extent to the optical light (see the left panel of 
Fig.\ref{fig:hstiracoverlay}). Because of the relative lack of  
8\,\micron\ emission in the outer disk, the flux ratio map  
(Fig.\ref{fig:iraccolor}) is very noisy there. Examination of the right panel 
Fig.\ref{fig:hstiracoverlay} shows that 
the 4.5\micron\ peak coincides with the bright blue star-forming nucleus 
of \mrk\ rather than with the red dust patch to the northwest 
\citep[see Fig. 4a of][]{thuan96}. 
This suggests that the MIR emission is mainly associated with 
the source responsible for the very red {\it nuclear} \jk\ color.  
Fig. \ref{fig:iraccolor} shows that the 4.5/8 flux
ratio is relatively constant ($\sim$0.7) outside the nuclear region
(see \S\ref{sec:iracdata}), although
beyond $r$ $\sim$18\arcsec, the color map becomes too noisy to give reliable 
colors.
This can also be seen in the lower left panel of Fig. \ref{fig:profile} 
where, for radii larger than $\sim$ 6\arcsec, the radial color profile shows a
relatively constant
[4.5]$-$[8.0] color of 1.5, corresponding to a 4.5/8 flux ratio of $\sim$0.7. 
The apparent gradient in the circumnuclear region at the radius of 
$\sim$ 4\arcsec\  
is due to the mismatch of the diffraction-limited IRAC point-response functions
at 4.5 \mic\ and 8 \mic.
Because the 4.5/8 color of stellar populations is $\sim2.5$,
roughly independently of age and metallicity \citep{sb99},
it is clear that the infrared emission of \mrk\ is dominated by the ISM,
and not by stars. 

If we interpret the 8\,\micron\ emission in the 
usual way and attribute it primarily to PAHs, then 
the 4.5/8 ratio $\sim$0.7 
implies a contribution from PAHs in the ISM around the 
nuclear star-forming region of \mrk.
This interpretation is indeed supported by the detection of
strong PAH features in the IRS spectrum discussed in the next section.
These PAHs 
are probably located in the outer low-density
region with 400 pc $\la$ $r$ $\la$ 1000 pc, since they   
are likely destroyed by the strong UV radiation field in the
nuclear region. 
However, our observations do not have sufficient spatial resolution
to verify this.

\section{Spectroscopic results}

\subsection{PAH features}

The IRS spectra (Fig. \ref{fig:irs}) show clearly PAH features at 5.7, 6.2, 7.7, 8.6, 11.2 and 12.8
\micron. The flux, the equivalent width EW and the full width at half maximum 
FWHM  
of each PAH feature have been derived by fitting Lorentzian profiles 
(Gaussian profiles do not give good fits) with 
the SPLOT task in IRAF. They 
are given in Table \ref{tab:pahs}. The continuum 
was linearly interpolated in two sections, one
from 6-9\,\micron, and the other from 10.5-13.9\,\micron. Standard deviations 
of repeated measurements are given in parentheses. 
The relative strengths of the three main PAH emission features can
be used as diagnostics to
identify PAH sizes and infer the neutral-to-ionized gas ratio
\citep{draine01}.
According to the models of \citet{draine01}, the 
observed PAH flux ratios  $f_{11.2}/f_{7.7}$ equal to 0.40 and 
$f_{6.2}/f_{7.7}$ equal to 0.27 indicate that  
the PAHs in \mrk\ are predominantly neutral and small, containing a 
few hundred carbon atoms, similar to the 
normal galaxies studied by \citet{helou00}. In other words, just as \hh, 
\mrk\ lies in 
the ``normal galaxy'' region in the PAH diagnostic diagram rather than in the
starburst galaxy region which includes objects such as M\,82 and NGC\,253,
that are dominated by ionized PAHs.
This would imply, at face value, that the PAHs in \mrk\ arise 
mainly from the normal cold and neutral ISM surrounding the nuclear region, 
but not from the star-forming region itself. 
However,
the PAH EWs are generally high, $\sim 3.6$\mic\ for the 7.7\,\micron\ line, which is 
more typical of starburst galaxies \citep{brandl04} than of
BCDs. \citet{wu06} found that the EW of the 7.7\,\micron\ line of all 
the BCDs in their sample to be less than 1 \micron, considerably smaller 
than the value of 3.6 \micron\ in \mrk. 
For the metallicity of \mrk, 
the EWs of its 6.2 and 11.2 \micron\ PAH features exceed by more  
one order of magnitude the values given by the mean EW (PAH) -- metallicity 
relations given by \citet{wu06} for their BCD sample.  

Just as for \hh, 
we find that the PAH emission features in \mrk\ are relatively narrow.
They are narrower than those predicted by 
the model for the
diffuse Galactic ISM at high latitudes \citep{li01}. While the 
6.2 \micron\ and 7.7 \micron\ features have FWHMs similar to those of 
the starburst galaxies M 82 and NGC 253, the 8.6 \micron\ feature is 
about half as wide while the 11.3 \micron\ feature is twice as wide. The best 
matches to the 
FWHMs of the PAH features in \mrk\ appear to be those of 
the reflection nebulae region NGC\,2023. Thus, neither the diffuse 
Galactic ISM model nor the starburst model of \citet{li01} and \citet{draine01} 
can account fully for all the 
observed properties of the PAHs in \mrk.  Even though the IRS spectrum of 
\mrk\ may contain a conspicuous
ISM component, Galactic ISM spectra are not necessarily good templates
for all PAH emission in this object.  
This suggests that there are many factors that determine the properties of 
PAH features (e.g., chemical abundance, ISM energetics, etc.) 
which are not yet taken into account fully in the models.
A more definitive observational 
test of the models, with a fuller exploration of the 
parameter space, will be possible with our larger sample of  
23 BCDs.

\subsection{Infrared Fine-Structure Lines \label{sec:lines}}

The spectrum in Fig. \ref{fig:irs} 
also shows several fine-structure lines. The 
\siv $\lambda$10.51, \neii $\lambda$12.81,       
\neiii $\lambda$15.55, \siii $\lambda$18.71, 33.58, and
 \oiv $\lambda$25.89 \micron\ are seen.
In the same way as for the PAH features, we used
SPLOT to fit the IR fine-structure emission lines and obtain
fluxes and EWs.
A deblending procedure was adopted to accurately measure
the emission lines at wavelengths near PAH features
(e.g., \neii).
The lines were best fit with Gaussian profiles.
A single continuum was linearly interpolated from 10.5 to $\sim15$\,\micron.
For the longer wavelengths, the local continuum was fit by linearly
interpolating over adjacent line-free regions. 
The continuum 1$\sigma$ uncertainty in the SH module is
$\simeq0.18\times10^{-17}$\,W\,m$^{-2}$, and
$\simeq0.11\times10^{-17}$\,W\,m$^{-2}$ in LH.
The fluxes and other parameters resulting from the fits
are given in Table \ref{tab:lines}.

Several fairly high-ionization lines are detected in \mrk.
The \oiv$\lambda$25.89\,\micron\ line with an ionization potential of 
54.9\,eV, just beyond the \heii\ edge at 54.4\,eV, is faint, but present
at the 4$\sigma$ level.
A blown-up view of the spectral region around the \oiv\ line is shown
in Fig. \ref{fig:oiv}.
However its presence raises a puzzle.
As discussed before, the optical spectrum of \mrk\ does not show 
the presence of the high-ionization \fev\ $\lambda$4227 emission line
which also has an ionization potential of 54.4\,eV.
The \heii\ $\lambda$4686 emission line, also with an 
ionization potential of 54.4\,eV, is seen, but its broad width suggests that 
it originates in WR stellar winds rather than in the ionized 
interstellar gas.  
This implies that the ionizing radiation in the optical line-emitting 
region is less hard than that in 
the MIR line-emitting region. 

We can have some idea of the hardness of the 
radiation in the MIR-emitting region by using 
line ratios of different ionic species of the same element,
such as \neiii\ and \neii, since these are sensitive to the shape of the 
spectrum of the ionizing radiation field. 
The ratio $F$(\neiii)/$F$(\neii) is 2.8, placing \mrk\
near the high-excitation end of the starbursts studied by
\citet{verma03}, similarly to NGC\,5253 and \hh. 
Because PAHs tend to be depleted in hard radiation fields,
their EW would be expected to be negatively correlated with 
the \neiii/\neii\ line ratio. 
Such a trend was indeed found by \citet{wu06} for their BCD sample. 
The anticorrelation between the PAH strength and the hardness and 
luminosity of the UV radiation field was also discussed 
by \citet{b06} in the case the irregular galaxy NGC 5253 and by 
\citet{l07} in the case of the massive young cluster NGC 3603.   
However, as discussed before, compared to the \citet{wu06} correlation,  
the PAHs in \mrk\ have considerably higher EWs for their metallicity.
This may again be due to a significant
PAH component from the general ISM in \mrk\ as compared to the BCDs studied 
by \citet{wu06}, as 
the region of \mrk\ covered by the IRS slit is relatively large, 
about 700\,pc.
The large EWs may also be the results of 
different measurement techniques, as \citet{wu06}
fit a spline to the local underlying continuum, while we linearly interpolate over
two wide sections of continua.

The long-wavelength IRS spectrum (Fig. \ref{fig:irs}) 
shows an emission feature at $\lambda$29.84\,\mic,
the wavelength of
one of the strongest \water\ lines in this spectral region. 
Such a detection would not perhaps be surprising,
as water can form in shock-heated gas associated with outflows
\citep[e.g.,][and references therein]{melnick08}.
However, we would also expect to detect a wealth of \htwo\ lines
which are not clearly seen in our spectrum (Fig. \ref{fig:irs}).
Hence, it is not clear whether this is a real feature or not. 
We will investigate the frequency of water emission
in a future paper on our larger BCD sample.



\subsection{The Infrared Spectral Energy Distribution}

We combine the available photometric and spectroscopic data to 
construct the IR spectral energy distribution (SED) in Fig. \ref{fig:sed}.
We have used the {\it total} emission from \mrk, integrated over
the entire galaxy including the disk.
The 8\,\micron\ IRAC point which is higher than the 
corresponding IRS point suggests that a significant part of the PAH emission
in the disk lies in a region more extended than the one 
covered by the IRS SL slit. 
The coincidence of the MIPS 24\,\micron\ point with the IRS spectral one
lends confidence to our data reduction procedures, including background 
subtraction and flux calibration.

It can be seen that the SED has a broad peak from 80
to 160\,\micron, implying that the dust in the disk of \mrk\ must be
rather cool, $\la$20\,K. 
Indeed, this broad peak toward longer wavelengths is almost certainly due
to the ``mixed morphology'' of \mrk. 
Up to $\la$70\,\micron, \mrk\ is
virtually a point source, 
dominated by 
the compact nuclear star-forming region;
but at longer wavelengths, the disk dominates with its contribution of cooler
dust.
Hence, the SED does not show a well-defined spectral peak, 
but rather a smeared-out
one, due to the contributions of dust at different temperatures from different
regions.


We have compared in Fig. \ref{fig:sed} the SED of \mrk\ with 
model SEDs of starburst galaxies such as  Arp\,220 and M\,82,  
taken from \citet{silva98}. 
These models, normalized to continuum emission between 20 and 30\,\micron,
clearly do not fit the SED of \mrk\ \citep[see also the discussion of][]{hunt05}:
\mrk\ shows no silicate absorption feature at $\sim$10\,\micron, 
relatively little PAH emission compared to a luminous solar-metallicity 
starburst galaxy, and a very different MIR continuum slope.
Moreover, in \mrk,
the unabsorbed emission that emerges at NIR and optical wavelengths
is greater relative to the dust emission, when compared  
to the prototypical starburst galaxies M\,82 and Arp\,220.
This is most probably due to a lower extinction and thus to a lower fraction
of dust reprocessing, but could also be due to differing dust content
relative to the stars.

We have also compared the SED of \mrk\ with that of the Seyfert 1 galaxy 
NGC\,4151 \citep{buchanan06},
 normalized over the 9 -15 \mic\ region (Fig. \ref{fig:agn}). 
While this SED gives low residuals in the normalization zone, il fails to 
fit the SED for wavelengths longer than 15 \mic. Clearly, dust reprocessing of 
the  H {\sc ii} region radiation dominates the SED of \mrk\ at longer wavelengths.
The surrounding cooler ISM also contributes significantly, resulting in 
a roughly constant (in flux density) SED from 70 to 160 \mic.   
     
 In summary, like most other star-forming 
galaxies \citep[see the discussion of the BCD 
\hh\ in ][]{hunt06}, \mrk\ is a composite entity in the IR; 
different components and morphologies are seen 
at different wavelengths. In \mrk, 
at $JHK$ we see the clear signature of a very red nucleus,
and the evolved extended underlying stellar population in the disk;
similarly, at 4.5\,\micron\ we see a nuclear hot dust continuum 
and the evolved stars in the disk; 
at 8\,\micron, we see mostly
extended PAH emission in the general ISM; 
at 24 and 70\,\micron, we see unresolved emission from warm dust 
associated with the nucleus; and  
at 160\,\micron\ cooler extended dust emission in the disk.

\section{CLOUDY modeling of the emission-line fluxes \label{cloudy}} 


\subsection{The narrow and broad line components of \mrk}

Thanks to a higher   
spectral resolution than that of the IRS spectrum,   
it can be seen clearly from the MMT optical spectrum 
(Fig. \ref{fig:sp}) that some lines 
have two components: a narrow component at high intensity levels 
and a broad component at low intensity levels. We have used SPLOT to 
deconvolve the profiles of these lines into narrow and broad components.
The deconvolution results are shown in Table \ref{tabdeconv}. It  
gives for both components the line flux of each detected 
line normalized to the \hb\ flux and 
its 
FWHM.
The narrow 
component of the [O {\sc iii}] $\lambda$4363 line is negligible. 
Three facts are to be noticed in Table \ref{tabdeconv}. First, the narrow 
line component has FWHMs varying between 2.6 and 3.5 \AA, 
comparable to the spectral resolution. It is thus not resolved, giving 
upper limits of  
gas velocities between 87 and 135 km s$^{-1}$. 
On the other hand, the broad 
line component has FWHMs varying between 6.1 and 8.5 \AA, 
corresponding to gas velocities between  
235 and 330 km s$^{-1}$. Second, the ratio of the auroral 
[O {\sc iii}] $\lambda$4363 line to that of the nebular 
[O {\sc iii}] $\lambda$4959 line for the 
broad component, equal to $\sim$ 0.45, is unusually high compared to its  
usual value of $\sim$ 0.1 for an electron temperature of 20 000 K, 
suggesting collisional deexcitation of the 
[O {\sc iii}] $\lambda$4959 line.
 Third, the intensities for the He {\sc i} lines relative to \hb\
are different for the narrow and broad line components.  
For the 
He {\sc i} $\lambda$3889 line, the relative intensity of the broad component 
is smaller  than that of the narrow component by a factor of 1.4, while for the  
He {\sc i} $\lambda$4471 line, the reverse is true, the broad component 
being more intense by a factor of 2. 

\citet{thuan96} have suggested that these observational features 
can be understood if the optical lines arise 
in two main regions: (1) the broad lines in a very dense inner region 
($r$ $\la$ 100 pc) 
with large gas 
mass motions, most likely powered by stellar winds of the 
Wolf-Rayet stars, the presence of which is clearly indicated by the 
WR bump in the MMT spectrum (Fig. \ref{fig:sp})  
, and (2) the narrow lines in a considerably less dense outer 
region ($r$ $\ga$ several hundred pc), 
with a density more comparable to those in normal \hii\ regions, 
and smaller mass motions. 
In this type of model, the helium lines would arise in the dense inner 
region: the intensity of the He {\sc i} $\lambda$4471 emission line would 
be enhanced because of collisional excitation, while the intensity of 
the He {\sc i} $\lambda$3889 line which is  
sensitive to optical depth effects, would be decreased because of  
the higher optical depth.
\citet{thuan96} were indeed able to construct an inhomogeneous 
two-zone CLOUDY model which 
accounts well for the observed line intensities of the UV and 
optical lines. Their best model has an inner region with  
a central electron density $N_e$ equal to 4.5 $\times$ 10$^6$ cm$^{-3}$
and a density decreasing with radius $r$ as $r^{-2}$ out to a radius of 100 pc, 
and an outer region with a constant density 
$N_e$ = 450 cm$^{-3}$ for radii between 100 pc and 
320 pc.  

In addition to the presence of distinct broad and narrow line regions, 
there is also a trend of increasing line width
with ionization potential, similar to Seyfert galaxies
\citep{derobertis84,whittle85}.
However the gas clouds motions and densities in the broad-line region  
are not as large as in the case of an AGN. Also, 
the UV and optical spectra show no evidence of the very hard ionizing 
radiation that is usually present in an AGN. But, as discussed 
before, the detection of the 
[O {\sc iv}] $\lambda$25.9\mic\ emission line in the IRS spectrum 
(Fig. \ref{fig:irs}) does imply the presence of hard EUV radiation in the 
MIR-emitting region. However, hard ionizing radiation, as seen below, 
can be produced by shocks or by WNE-w stars, not necessarily by an AGN.


\subsection{A hidden H {\sc ii} region}

We now use CLOUDY to check 
whether the two-zone 
model constructed by 
\citet{thuan96} to account for the 
UV and optical line fluxes observed by {\sl HST}
 can also explain the MIR line fluxes observed by \spitzer. 

The parameters of the 
CLOUDY model used here  
are very similar to those of model 1 of \citet{thuan96} 
(the one with low nitrogen abundance). The slight differences are due to 
the updating of the CLOUDY code with new atomic parameters. 
However, since the {\sl Spitzer} and MMT spectra have been obtained through 
significantly larger apertures than the {\sl HST} aperture, 
we have  
increased the radius of the modeled \hii\ region to $\sim$ 580 pc in order
to take into account the total H$\alpha$ emission of \mrk. This emission  
extends over an angular radius of $\sim$ 5\arcsec\ ($\sim$520\,pc)
\citep{gildepaz03}, and is covered entirely by the LH IRS slits and
mostly by the SH IRS slits.
The model gives a \hb\ luminosity $L$(\hb) = 2.34$\times$10$^{40}$ erg s$^{-1}$,
corresponding to a 
number of ionizing photons per second, log $Q$(H) = 52.75.
At the distance of 21.7 Mpc, this luminosity corresponds to a H$\beta$ flux
$I$(H$\beta$) = 4.18$\times$10$^{-16}$ W m$^{-2}$. This flux is 
very similar to
the H$\beta$ flux of 4.40$\times$10$^{-16}$ W m$^{-2}$ derived from the 
extinction-corrected H$\alpha$ flux measured for Mrk 996 by \citet{gildepaz03},
adopting an extinction coefficient $C$(H$\beta$) = 0.53 and a 
H$\alpha$/H$\beta$ ratio of 2.8.
As for the run of the number density with radius $r$, 
we have also adopted for the inner part 
(1 pc $\leq$ $r$ $\leq$ 100 pc) a $r$$^{-2}$ law, with a maximum 
log $N_e$ = 6.66
at the inner boundary ($r$ = 1 pc), and a constant log $N_e$ = 2.645 for 
$r\geq$ 100 pc. The filling factor, 
log $f$ = $-$4.58, characteristic 
of a very clumpy interstellar medium, is the same as
in \citet{thuan96}. Hereafter, this model will be referred to as Model I.

\subsubsection{Comparison of observed and modeled optical line fluxes}

First, we check that Model I is indeed 
able to reproduce the fluxes of the optical
emission lines obtained from the MMT spectrum. Table \ref{tabcomparopt}
lists the extinction-corrected observed fluxes 
and Model I fluxes of several strong 
optical lines along with their ratios. It is seen that  
the observed high-ionization [O {\sc iii}] and [Ne {\sc iii}] emission
lines are well reproduced by the model, but that the observed fluxes of 
of the lower ionization [O {\sc ii}] $\lambda$3727 and H$\beta$ emission 
lines are significantly 
smaller (by factors of 2 and 1.7 respectively) 
than the modeled ones. Evidently, the difference is due to
aperture effects because the aperture of the MMT spectrum does not
cover all the emission from the \hii\ region, and 
the lower-ionization lines are produced further away from the 
nucleus than the high-ionization ones. Therefore, we have also calculated  
Model Ia, which has the same parameters as Model I,
except that the radius of the modeled 
\hii\ region is equal to 410 pc instead 
of 580 pc. This radius was chosen so that the modeled H$\beta$ flux matches 
the observed H$\beta$ flux within the MMT aperture. Compared to  
Model I fluxes, the fluxes of the
high-ionization lines in Model Ia remain unchanged, while the flux of the
[O {\sc ii}] $\lambda$3727 emission line is considerably lower, bringing it 
into good agreement with the observed one, as 
shown in the last column of Table \ref{tabcomparopt}. Thus, we conclude that
CLOUDY models with the parameters adopted above do indeed  
reproduce well the observed fluxes
of the optical lines.

\subsubsection{Comparison of observed and modeled MIR line fluxes}

We next compare the observed MIR emission line fluxes with those 
predicted by Model I.   
Examination of columns 2 and 3 of 
Table \ref{tabcompar} shows that, despite similar 
observed and modeled fluxes of the optical lines, 
Model I predicts significantly
lower MIR line fluxes, as compared to the observed ones, 
except for [Ne {\sc ii}]
$\lambda$12.8 $\mu$m. There could be at least two explanations of 
this discrepancy: 1) aperture effects and 2)
the presence of an additional obscured \hii\ region that is not seen in
the optical range, but that contributes to the MIR emission.

We consider first the possible role of aperture effects. The angular radius of
the optical \hii\ region of $\sim$ 5\arcsec\ is significantly greater 
than half of the slit width used 
in the optical observations. It is greater than or
comparable to half of the slit width used for the IRS spectra. Therefore,
aperture effects may be important.  
In particular, these effects may play a role 
for the hydrogen lines and the low-ionization forbidden optical  
[O {\sc ii}] $\lambda$3727 and the MIR [Ne {\sc ii}] $\lambda$12.8\micron\
lines because a significant fraction of emission in 
these lines comes from the outer zones of the \hii\ region. For 
higher-ionization lines, the aperture effects should be smaller 
because those
lines arise in the inner zones of the \hii\ region. To check how aperture
effects may influence the emission line fluxes, we show in Figure 
\ref{fig:aperture}
the fluxes $F$ predicted by CLOUDY 
of different optical and MIR lines emitted by a \hii\ sphere with
a varying radius $r$, keeping all other parameters equal to those 
of Model I. All fluxes are normalized to the
fluxes $F_{max}$ from the sphere with the 580 pc outer radius of the \hii\ 
region in Model I. We have also shown by vertical lines the ``equivalent''  
radii of the  
different apertures used in the optical and MIR observations, 
defined as $\sqrt{S/\pi}$, where $S$ is the area of the aperture. We
also mark the location of the outer radius of the \hii\ 
region in Model I. The 
$F/F_{max}$ ratios cannot be used directly for aperture corrections since they
are derived for spheres, while the observed emission comes 
from a cylindric column with equivalent radius $r$. They
provide however upper limits to the aperture corrections and  
Fig. \ref{fig:aperture} can be used for a qualitative analysis.
It is seen from that figure that 
aperture corrections are important for some optical
lines, especially the [O {\sc ii}] $\lambda$3727 line, as confirmed  
by comparison of the {\sl HST} and MMT fluxes: the MMT flux 
(Table \ref{tabint}) 
is 1.5 times larger than the {\sl HST} flux \citep{thuan96}.
The {\sl HST} and MMT H$\beta$ fluxes are respectively $\sim$ 3 and 
2 times lower 
than the one inferred from
the H$\alpha$ flux by \citet{gildepaz03}. On the other hand, the aperture
correction is negligible for higher-ionization optical lines, such as
[O {\sc iii}] $\lambda$5007. The aperture corrections for the MIR lines are
significantly lower because of the larger apertures used. 
Fig. \ref{fig:aperture} shows that the  
IRS SH aperture includes nearly all the \hii\ region emission and the 
IRS LH aperture includes all of it. Thus, aperture effects cannot
account for 
the large differences between the observed and modeled fluxes of the 
MIR lines, except perhaps for the 
[Ne {\sc ii}]$\lambda$12.8$\mu$m emission line which has the largest  
aperture correction. 
We conclude that, to account for the MIR emission, 
we need to postulate the existence of 
a hidden H {\sc ii} region in
Mrk 996 that contributes to the MIR emission, but not to the optical emission.

We have found that the best CLOUDY model of this obscured 
\hii\ region (hereafter Model II) is characterized by  
the following 
parameters: a number of ionizing photons log $Q$(H) = 52.425 -- $\sim$ 2 times
lower than for Model I --, an effective temperature $T_{eff}$ = 50000 K, 
-- slightly higher than in Model I, 
suggesting a younger H {\sc ii} region --,
a constant number density of 300 cm$^{-3}$, a filling factor log $f$ = --2, 
an H {\sc ii} region radius of $\sim$ 100 pc and 
the same chemical composition as in Model I. 
The predicted
fluxes of the MIR lines for Model II are shown in Table \ref{tabcompar}. It is
seen that, except for the
[Ne {\sc ii}]$\lambda$12.8$\mu$m line, 
the predicted fluxes of the MIR lines in Model II (the hidden 
\hii\ region) are all significantly higher
than those in Model I (the visible \hii\ region), despite a number of 
ionizing photons which is $\sim$2 times lower.
The main reason is that, in Model I, the infrared fine-structure line
fluxes are strongly reduced by collisional deexcitation of the upper
levels in the very dense central part of the visible H {\sc ii} region. 
At the critical densities, the line ratios will change as some lines will 
be suppressed less strongly as compared to other lines that will be 
suppressed more strongly.
The critical number densities for collisional 
deexcitation of the detected MIR lines are in the range $\sim$10$^3$ --
10$^5$ cm$^{-3}$ \citep{m02}, lower than the central number density.     
Collisional deexcitation is however not important for the 
[Ne {\sc ii}]$\lambda$12.8$\mu$m line because it has a very high
critical density of 6.1$\times$10$^{5}$ cm$^{-3}$ and in addition it arises 
in the outer less dense zones of the \hii\ region in Model I.
Adding the line fluxes from Models I and II (columns 3 and 4 of 
Table \ref{tabcompar}), we see that the modeled and observed fluxes 
of the MIR lines are in general good agreement, their ratios (column 8 of 
Table \ref{tabcompar}) being close to unity. How much dust extinction would 
it take to hide the \hii\ region of Model II ? If we set the upper limit 
of its H$\beta$ flux to be 10\% that of the visible region, 
then its $A_V$ would be $\ga$ 4 magnitudes.



\subsection{A hidden source of hard ionizing radiation}

There is, however, one glaring discrepancy in column 8 of 
Table \ref{tabcompar}:  
the model does not predict any  
[O {\sc iv}] $\lambda$25.9\mic\ emission, while this line is 
clearly detected in the IRS spectrum. The ionization potential of 
[O {\sc iv}] is 54.9 eV, just beyond the He {\sc ii} edge at 54.4 eV.
As discussed before, 
we have searched the optical spectrum of Mrk 996 for the high-ionization 
[Fe {\sc v}] $\lambda$4227 and He {\sc ii} $\lambda$4686 nebular emission 
lines which both have an ionization potential of 54.4 eV, but none were seen.
This implies that the hard ionizing radiation comes entirely from a  
region that is optically hidden, which is not necessarily spatially 
coincident with the optically hidden H {\sc ii} region discussed before.  
   
 
Mrk 996 is not the only BCD to reveal harder ionizing radiation
in the MIR than the optical. The same phenomenon occurs in 
the BCD Haro 3, the most metal-rich [12+log(O/H)= 8.32] BCD of 
our \spitzer\ sample. 
Haro 3 also shows a well detected [O {\sc iv}] $\lambda$25.9\mic\ 
emission line in its IRS spectrum, while 
the [Fe {\sc v}] $\lambda$4227 and He {\sc ii} $\lambda$4686 
nebular emission lines are conspicuously absent
from its optical spectrum \citep{hunt06}.
We will have to wait for the analysis of our whole \spitzer\ BCD sample to tell 
whether that is a general property of BCDs, or whether this occurs only 
in exceptional cases. 

 What is the origin of  
the hard ionizing radiation beyond 4 Ryd  in \mrk? 
Can stellar radiation be responsible? We have run Costar models by 
\citet{schaerer97} with stars having 
the highest effective temperature (T$_{eff}$ = 53 000 K)
and the hardest radiation possible (corresponding to the hottest O3 stars).  
We failed to reproduce the observed intensity of the 
[O {\sc iv}] $\lambda$25.9\mic\ by a factor of several. 
We have then considered  Wolf-Rayet stars of type WNE-w. According to 
calculations by \citet{crowther99}, models for WNE-w stars (early 
nitrogen Wolf-Rayet stars with weak lines) show a strong 
ionizing flux above 4 Rydberg ($\lambda$ $\leq$ 228 \AA), in contrast 
to the WCE and WNL stars which show negligible fluxes above 3-4 Rydberg.
These stars have transparent winds, and therefore their 
$Q$($>$ 4 Ryd)/$Q$ ($>$ 1 Ryd), where $Q$ is the ionizing flux, is some 100 times 
higher than the same ratio predicted for the hottest O stars 
by the Costar models \citep{schaerer97}.  We have run CLOUDY models with 
the ionization spectrum of a WNE-w star \citep{crowther99}. We 
found that with a log $Q$(H) = 51.425, some 25 times less than that 
from the bright H {\sc ii} region, and some 10 times less than 
the $Q$(H) of the hidden H {\sc ii} region, the observed intensity of 
the [O {\sc iv}] line can be reproduced (Table \ref{tabcompar}). 
The fraction of WNE-w to O stars is several percent if the same  
$Q$(H) (log $Q$(H) $\sim$ 49) is assumed for single O and WNE-w stars.
This estimate may be wrong by a factor of 10 because of the uncertainty in 
the $Q$(H) of WNE-w stars. 
The WNE-w stellar  
spectrum does not produce much of the lower-ionization 
species, so that  
the agreement obtained previously between observations and models for the line 
intensities of the lower-ionization 
species is not destroyed. A weak [Ne {\sc v}] $\lambda$24.3\mic\ line 
is predicted by CLOUDY, with an intensity some 9 times weaker  
than that of the [O {\sc iv}] line intensity, consistent with its 
non-detection.
How plausible is the presence of such a WNE-w stellar population 
in the central star-forming region of \mrk? 
A ratio $N$(WNE)/$N$(O) of a few percent appears high because, at the metallicity 
of \mrk\, it is roughly equal to the number ratio of WR stars of all types to 
O stars \citep{guseva00}, and because WNE-w stars are rare among WR stars.
Moreover, from the equivalent width of 
H$\beta$ (106.9 \AA), we estimate the age of the bright visible H {\sc ii}
region to be 3-4 Myr. WNE stars are thought to result from mass transfer  
in close binary systems, so they dominate the stellar population at ages 
$\geq$ 10 Myr, when O stars and WR stars of other types are gone. Thus 
the visible  H {\sc ii} region is too young to contain WNE stars. 
If one wishes to invoke WNE stars as the source of hard radiation,
one would have to postulate that these stars are not visible in the optical
range. All things considered, we do not consider the WNE-w hypothesis very likely. 

We have also considered the case of a non-stellar source of ionizing 
radiation, that emitted by an accretion disk around an 
intermediate-mass black hole in the 
center of \mrk. Thus, we have also run CLOUDY models with an ionization 
spectrum in the 
form of a power-law: $f$($\nu$) $\propto$ $\nu^{-1}$.  We found that,
in order to reproduce the line intensity of the [O {\sc iv}] line,
the number of ionizing photons coming from the AGN has  
to be log Q (H) = 51.125, some 50 times smaller than that  
coming from the bright H {\sc ii} region. Again, the previous agreement between 
models and observations for 
the lower-ionization species is preserved.  
The results are also given in Table \ref{tabcompar}. Again, 
a weak [Ne {\sc v}] $\lambda$24.3\mic\ line is predicted, with an intensity 
some 3 times weaker than that of the [O {\sc iv}] line intensity. This 
line can be even weaker if the power-law spectrum has a slope steeper than $-1$. 
In any case, such a low-intensity line would be hard to detect. Because we
have to postulate that the AGN is completely invisible in the optical, we 
do not favor the AGN hypothesis as the most probable one.

Finally, we discuss a third possible source of hard 
ionizing radiation. \citet{izotov01}, \citet{izotov04} and 
\citet{thuanizotov05} have suggested that 
fast radiative shocks moving through a dense 
interstellar medium with $N_e$ $\sim$100 cm$^{-3}$ 
can be possible sources of photons with energy $\ga$54 eV,
and be responsible for the [Ne {\sc v}] $\lambda$ 
3426 (ionization potential of 7.1 ryd), [Fe {\sc v}] $\lambda$ 
4227 (ionization potential of 4 ryd), and He {\sc ii} $\lambda$ 4686 
(ionization potential of 4 ryd) emission they observed in some BCDs. 
The ionizing spectrum of such fast shocks has been computed 
by \citet{dopita96} for a gas of solar metallicity and various shock 
velocities. Recently, this work has been extended to environments 
with lower metallicities by M. G. Allen et al. 2004 (in preparation).  
Thus, we have also run CLOUDY models with an  
ionizing spectrum taken from the ``Mapping'' III Shock 
model library ($http://cdswww.u-strasbg.fr/~allen/mappings{_-}page1.html$),
and characterized by a metallicity equal to that of the Small Magellanic Cloud 
(closest to the metallicity of \mrk) and a shock velocity equal to 
250 km s$^{-1}$. This 
velocity corresponds to the FWHM of $\sim$ 7 \AA\ observed 
for the broad component of the 
emission lines in Table \ref{tabdeconv}. 
The CLOUDY results are given in Table \ref{tabcompar} for the shock model
with a number of ionizing photons log $Q$(H) = 51.425, the same as that
for the model with the WNE-w stars. It can 
be seen that the predicted line intensities in the shock model are 
very similar to those in the WNE-w model, and thus can account equally 
well for the intensity of the [O {\sc iv}] line. Again a weak [Ne {\sc v}] 
$\lambda$24.3\mic\ line is predicted which would be undetectable.        
We favor the shock hypothesis as the most plausible explanation 
for the [O {\sc iv}] line, as  
\mrk\ contains many WR stars with 
outflowing stellar winds and, likely, supernova remnants 
which will no doubt cause radiative shocks.

\section{Summary\label{sec:conclusions} }

We have acquired {\sl Spitzer} MIR, UKIRT NIR and MMT optical observations 
of the blue compact dwarf galaxy \mrk\ to study its gas, dust and stellar content.
 This BCD, with a metallicity of about 1/5 that of the Sun, 
has the peculiarity of 
possessing an extremely dense nuclear star-forming region: 
its central density is 
$\sim$ 10$^6$ cm$^{-3}$, some 4 orders of magnitude greater than the densities 
of normal \hii\ regions. 
We have obtained the following results:

\noindent
(1) The nucleus of \mrk\ is extremely red, with \jk\ = 1.8, and
\hk\ = 1.0, probably due to very hot dust with a temperature between 600 
and 1000 K. 
The optical spectrum of the BCD  shows the high-ionization lines to  
have both broad and narrow line components, 
and a trend of increasing line width with increasing ionization potential.  
     
\noindent
(2) The $VIJHK$ colors of the underlying exponential disk are roughly
consistent with the colors of a coeval stellar population with age $\ga$ 1 Gyr.  


\noindent
(3) Like most star-forming galaxies, \mrk\ is a composite entity in the IR.
We see extended photospheric emission from evolved stars,  
compact hot dust continuum 
coming from the nuclear star-forming region at 4.5\,\micron, 
hot dust continuum 
and extended PAH emission coming mainly from 
the surrounding less dense ISM at 8\,\micron,
compact small grain warm dust 
associated with 
the active star-forming nuclear region at 24\,\micron\ and 70\,\micron, 
and cooler extended dust 
emission associated with older stellar populations at 160\,\micron.  

\begin{figure}
\vbox{
\includegraphics[angle=0,width=0.8\linewidth,bb=18 165 590 543]{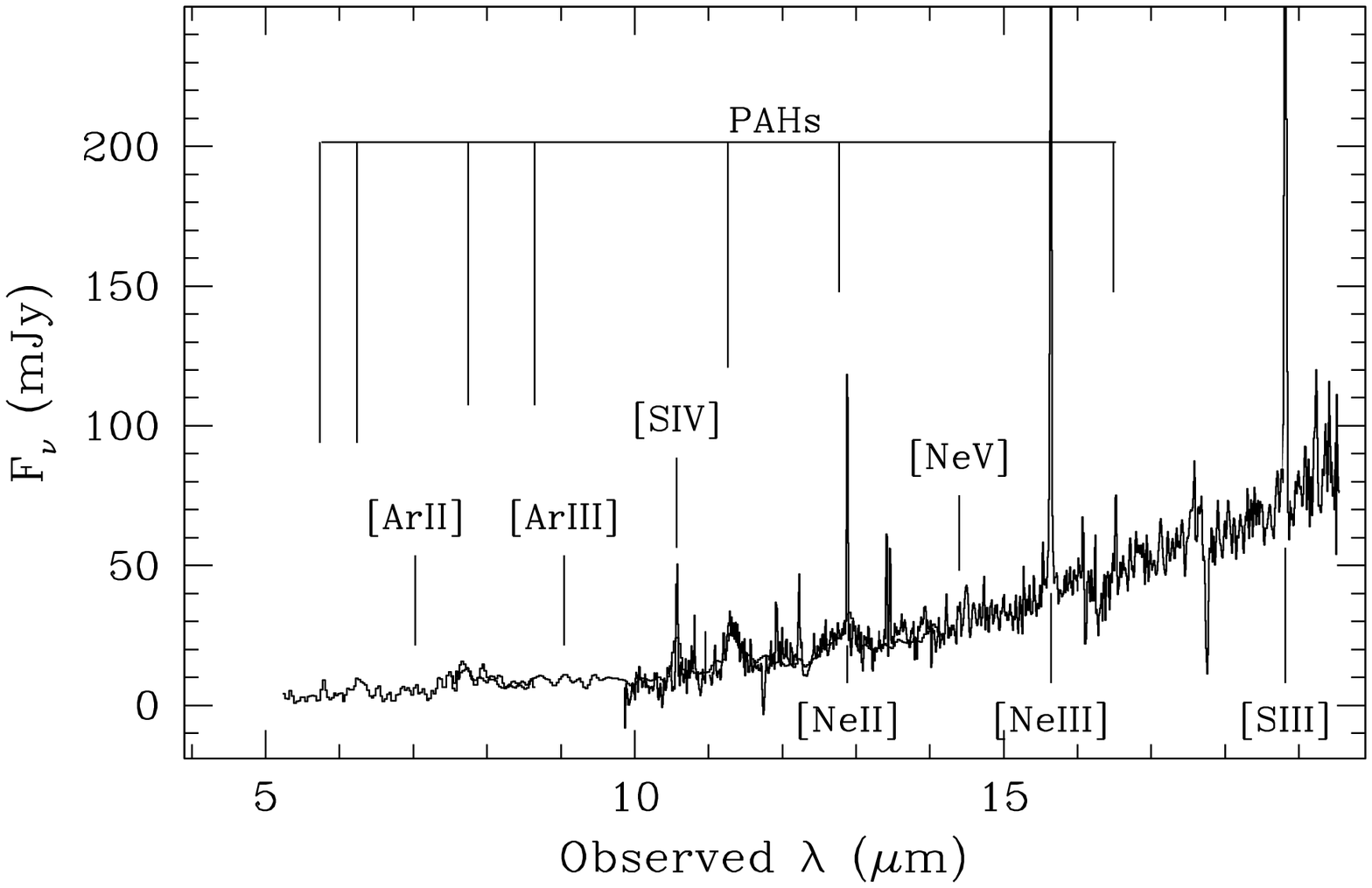}
\includegraphics[angle=0,width=0.8\linewidth,bb=18 165 590 543]{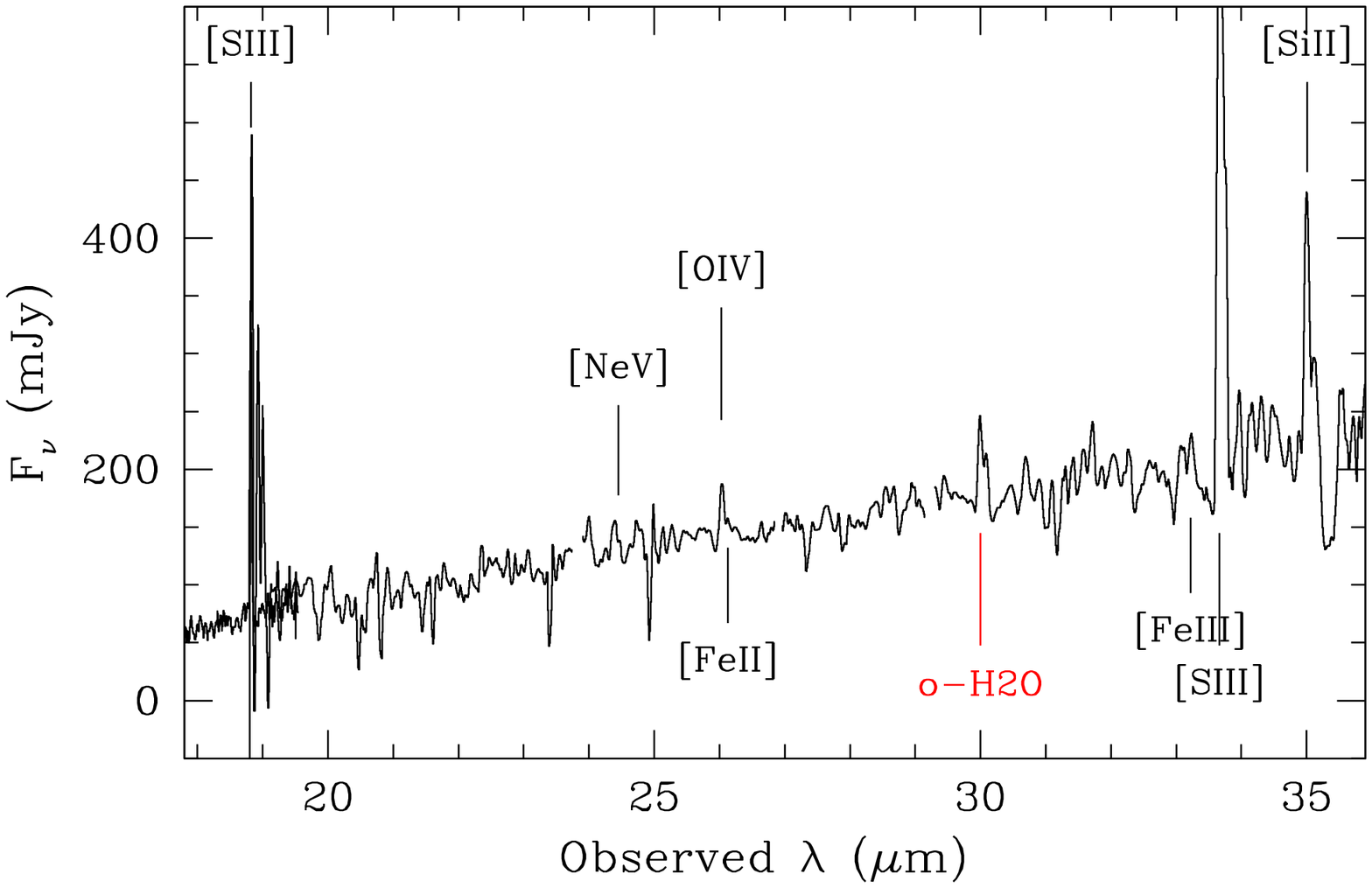}
}
\caption{The IRS spectrum of \mrk\ taken in the short-wavelength low-resolution mode
and in both high-resolution modes:
SL$+$SH spectra are shown in the upper panel, and the LH spectrum 
in the lower one. 
The PAH features at 5.7, 6.2, 7.7, 8.6, and 11.2 \micron\
are clearly detected, as are several fine-structure emission lines:
\siv, \neii, \neiii, \siii, \oiv.
\label{fig:irs}}
\end{figure}

\begin{figure}
\includegraphics[angle=0,width=0.8\linewidth,bb=18 165 590 543]{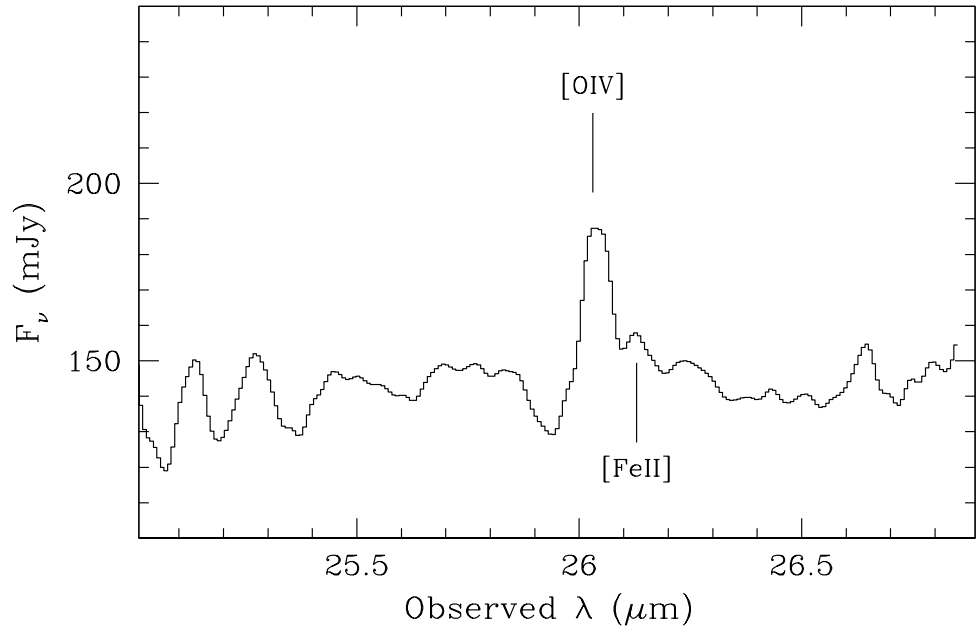}
\caption{A blow-up of the IRS spectrum of \mrk\ around the \oiv\ 25.89 \mic\ line.
\feii\ may be present, but is below the detection limit of the spectrum.
\label{fig:oiv}}
\end{figure}

\begin{figure}
\hbox{\includegraphics[angle=0,width=0.8\linewidth]{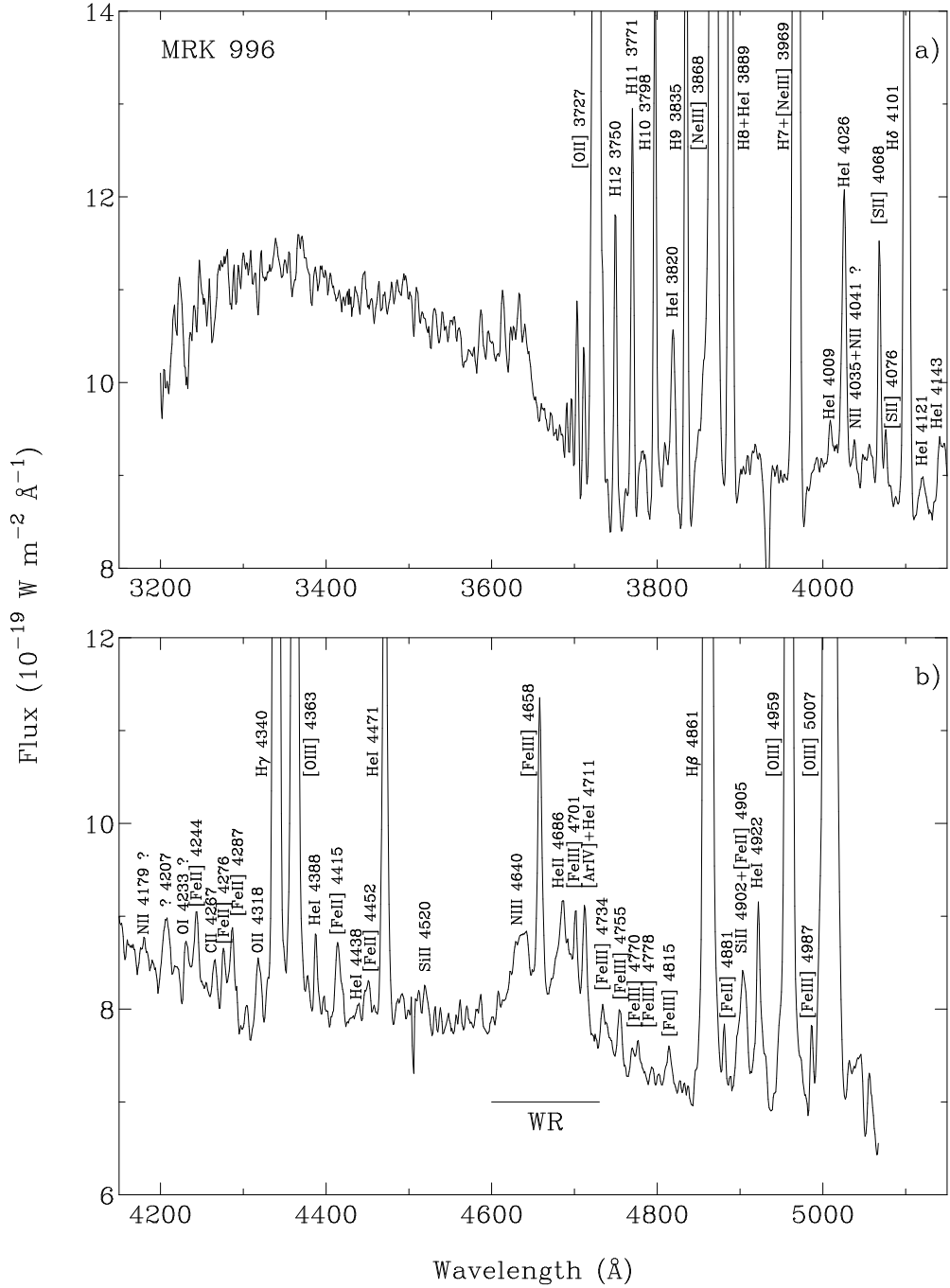} }
\caption{MMT spectrum of Mrk 996.
\label{fig:sp}}
\end{figure}

\begin{figure}
\vspace{0.2cm}
\hbox{
\includegraphics[angle=0,width=1.0\linewidth]{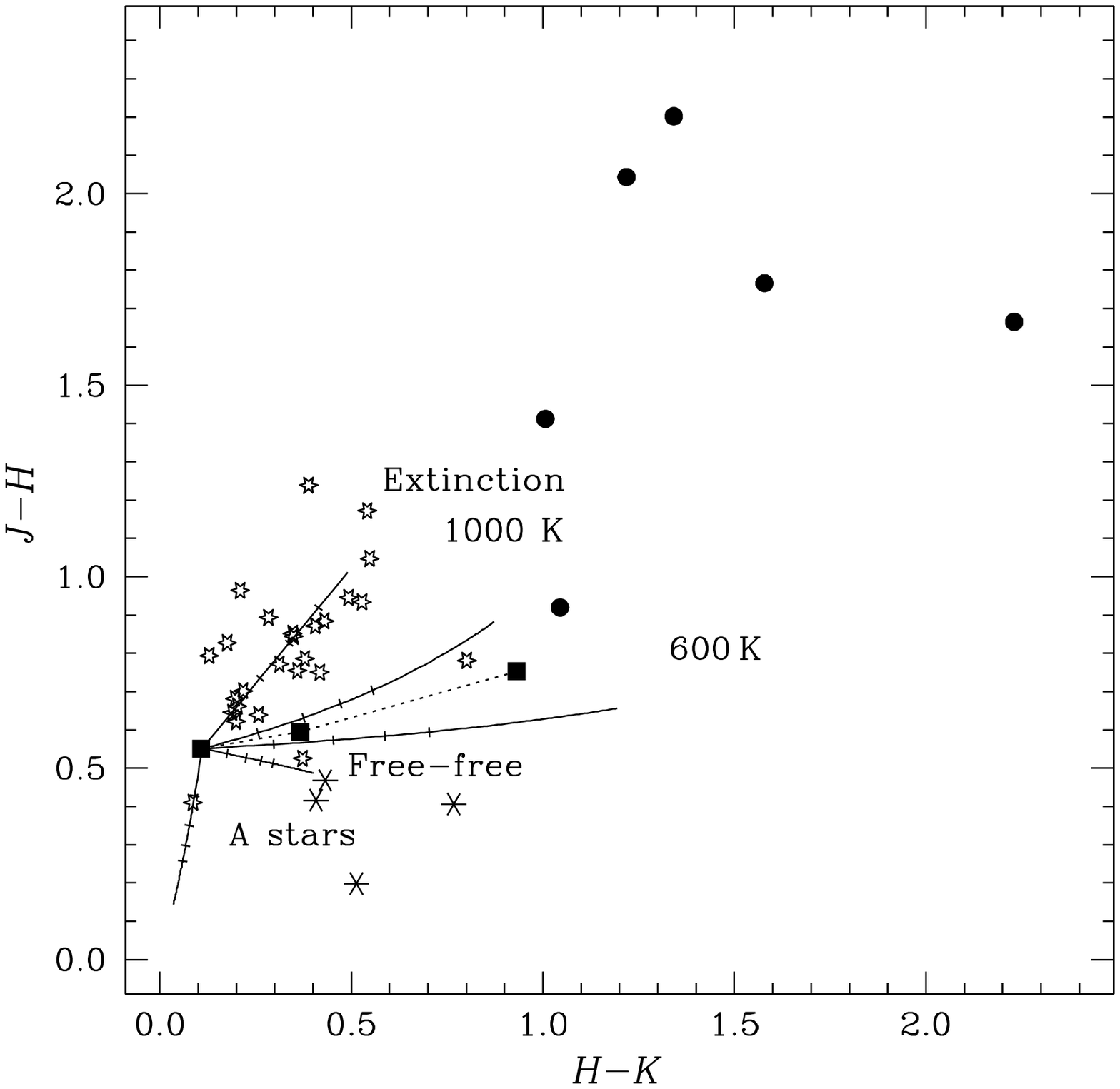} }
\caption{The (\jh,\hk) diagram. The colors of the nucleus of Mrk 996 
and of two locations outside of it (filled squares from right to left 
connected by a dotted line) 
  are compared to those of starburst galaxies (stars) 
\citep{hunt02}, blue compact 
dwarf galaxies (asterisks) \citep{hunt02} and Seyfert galaxies (filled 
circles) \citep{alonso01}.  Mixing curves show how the colors change
 when various 
physical components contribute in increasing amounts to the observed 
disk emission 
\citep[see][]{hg92}. From top to bottom, the curves show the 
influence of extinction, of hot dust at 600\,K, and at 1000\,K, and of
A stars;  
the end points of
the mixing curves correspond to equal $K$-band contributions from stars
 and from the various components.
The tick marks on the extinction line show magnitude
 increments in $A_V$ units.
\label{fig:jhk}}
\end{figure}

\begin{figure}
\hbox{\includegraphics[angle=0,width=1.0\linewidth]{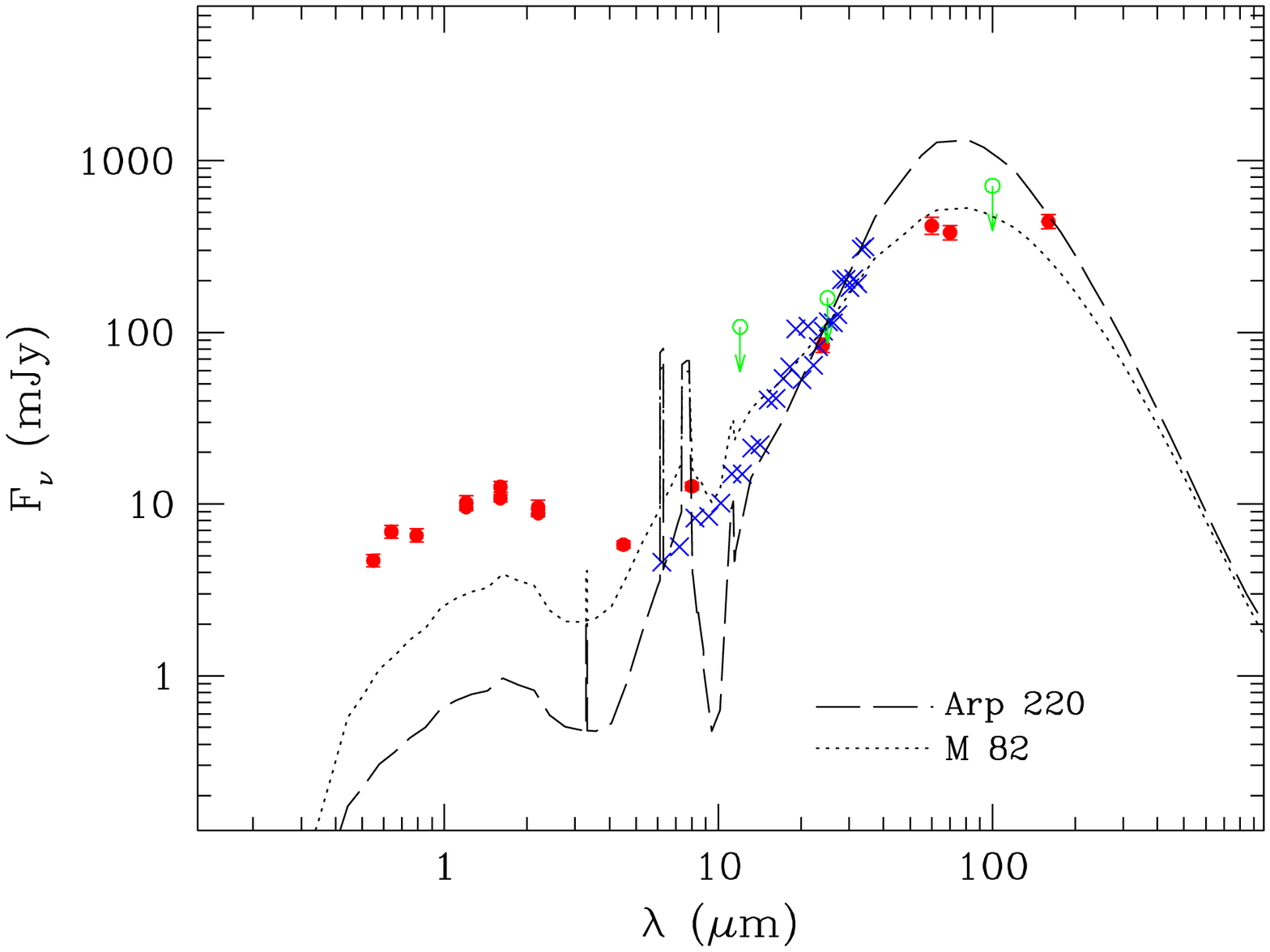} }
\caption{The SED of \mrk\ obtained by combining our UKIRT and \spitzer\
data with published data. 
The broadband data points are given by filled circles with error bars, 
upper limits
by open circles, and IRS spectral points by $\times$;
the IRS spectrum has been rebinned into 1\,\micron\ intervals.
The global optical fluxes are taken from \citet{thuan96}, assuming
colors within a 9\arcsec\ aperture centered on the nucleus.
Also shown are GRASIL \citep{silva98} models of Arp\,220 (as a long-dashed
curve) and M\,82 (dotted),
scaled to the MIR flux between 20 and 30\,\micron.
\label{fig:sed}}
\end{figure}

\begin{figure}
\hbox{\includegraphics[angle=0,width=1.0\linewidth]{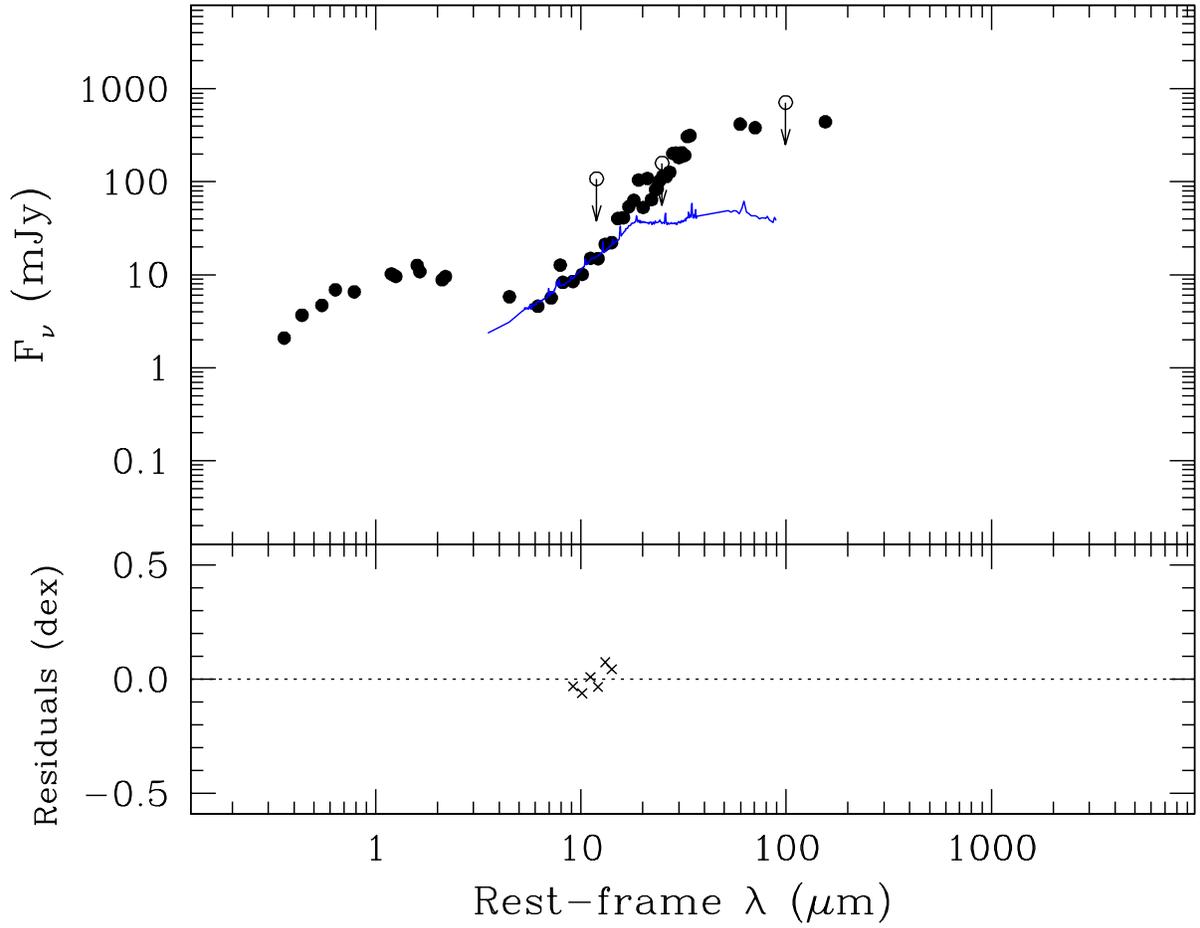} }
\caption{Comparison of the \mrk\ SED with that of the Seyfert I galaxy NGC\,4151
 \citep{buchanan06}.
\label{fig:agn}}
\end{figure}

\begin{figure}
\hbox{\includegraphics[angle=0,width=0.8\linewidth]{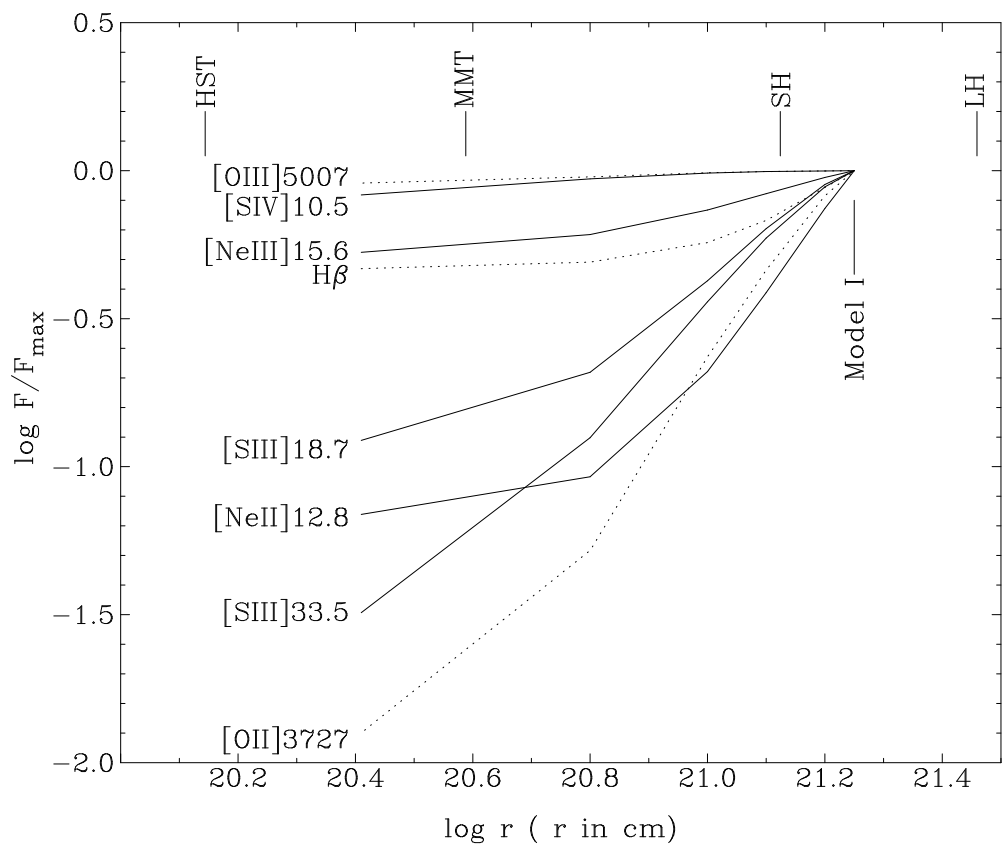} }
\caption{Fluxes $F$ of the infrared lines (solid lines) listed in Table 
 \ref{tab:photom} and of some optical
lines (dotted lines) computed 
with CLOUDY for a spherical high-density \hii\ region 
with radius $r$. They are normalized to $F_{max}$, 
the flux at log $r$ = 21.25, where
$r$ is in cm.
\label{fig:aperture}}
\end{figure}

\noindent
(4) The IRS spectrum (Fig. \ref{fig:irs}) 
shows strong Polycyclic Aromatic Hydrocarbon (PAH) molecular  
emission, with features clearly detected at 5.7, 6.2, 7.7, 8.6, 11.2 and 
12.7\,\micron. The PAHs in \mrk\ are predominantly neutral and 
small, similar to those found in normal spiral galaxies, suggesting that 
they reside in the general ISM and not in the star-forming region. The PAH 
emission features are relatively narrow and their equivalent widths are 
generally high for the metallicity of \mrk,  
exceeding by more than one order of magnitude 
the values given for the mean EW(PAH)-metallicity relation derived 
by previous investigators.
   
\noindent
(5) Gaseous nebular line emission is seen. The IRS spectrum 
shows several fine-structure forbidden lines, including
\siv\ $\lambda$10.51, \neii\ $\lambda$12.81,       
\neiii\ $\lambda$15.55, \siii\ $\lambda$18.71, 33.48 and 
\oiv\ $\lambda$25.90\,\micron. 

\noindent 
(6) We have used CLOUDY to model the line-emitting region.
To account for both the optical and MIR lines, two \hii\ regions are
required: a) a very dense \hii\ region that is seen in the optical
range (Model I) and b) an optically obscured \hii\ region ($A_V$ $\ga$ 
4 mag) with a constant number density of $\sim$ 300 cm$^{-3}$, 
typical of other \hii\ regions (Model II).
A two-zone model is required for the non-obscured \hii\ region (Model I): 
a) a very dense nuclear region where the broad 
optical line components arise; from a central value of $\sim$ 10$^6$ 
cm$^{-3}$, the density decreases with 
distance as $r^{-2}$ until $r\sim$ 100 pc; 
b) an outer zone for 100 pc $<r<$ 580 pc, with constant number 
density of $\sim$ 400 cm$^{-3}$ . The density gradient is 
probably caused by large scale gas mass motions, powered by the stellar winds 
of Wolf-Rayet stars.

\noindent
(7) The UV and optical spectra show no evidence of the very hard ionizing radiation seen in AGN. The IRS spectrum does show however 
the presence of a faint \oiv\ line at 25.89\,\micron, 
indicating 
the presence of radiation as hard as 54.9 eV in \mrk. 
This hard radiation is most likely due to fast radiative shocks in 
the ISM caused by outflowing stellar winds from WCE and WNL stars
and/or by supernova remnants. A  
hidden AGN, or a population of hidden 
Wolf-Rayet stars of type WNE-w are are less likely sources of hard ionizing
radiation. 
   

\acknowledgments

This work is based on observations made with the {\sl Spitzer}
Space Telescope, which is operated by the Jet Propulsion Laboratory, 
California Institute of Technology, under NASA contract 1407. 
The MMT time was available thanks to a grant from the Frank Levinson Fund 
of the Peninsula Community Foundation to the Astronomy Department of the 
University of Virginia.
We thank Jack Gallimore for providing us with the spectrum of NGC\,4151.
We are grateful to the UKIRT Time Allocation Panel for a generous time allocation.
Support for this work was provided by NASA through contract 1263707 
issued by JPL/Caltech. 
T.X.T. and Y.I.I. also acknowledge partial financial 
support from NSF grant AST 02-05785. L.K.H. and Y.I.I. are grateful to the 
hospitality of the Astronomy Department of the University of Virginia where part of 
this work was done.

\clearpage

\clearpage

\clearpage

\clearpage

\clearpage

\clearpage

\clearpage

\clearpage

\clearpage

\clearpage

\clearpage

\begin{deluxetable}{ccc}
\tablecaption{IRAC and MIPS total fluxes of \mrk\ with Other 
Published Photometry \label{tab:photom} }
\tablewidth{0pt}
\tablehead{
\colhead{Telescope/Instrument} & \colhead{Wavelength} & \colhead
{Total flux} \\
\colhead{} & \colhead{(\mic)} & \colhead{(mJy)} \\
}
\startdata
Spitzer/IRAC & 4.51 & 5.8 $\pm$ 0.29  \\
" & 7.98 & 12.7 $\pm$ 0.64  \\
Spitzer/MIPS & 23.7 & 84 $\pm$ 4 \\
" & 71.0 & 381 $\pm$ 10 \\
" & 156.0& 442 $\pm$ 12 \\
\hline
IRAS\tablenotemark{a} & 12 & $<$ 108 \\
"                     & 25 & $<$ 159 \\
"                     & 60 & 417 $\pm$ 50 \\
"                     & 100 & $<$ 713 \\
\hline
Optical\tablenotemark{b} & 0.36 &  2.1 $\pm$ 0.2 \\
"                        & 0.44 &  3.7 $\pm$ 0.3 \\
"                        & 0.55 &  4.7 $\pm$ 0.4 \\
"                        & 0.64 &  6.9 $\pm$ 0.6 \\
"                        & 0.79 &  6.6 $\pm$ 0.6 \\
\hline
IRCAM/UKIRT\tablenotemark{c}  & 1.26 &  9.6 $\pm$ 0.5 \\
"                       & 1.65 & 10.8 $\pm$ 0.5 \\
"                       & 2.12 &  8.9 $\pm$ 0.4 \\
\hline
2MASS\tablenotemark{a}	& 1.2 & 10.2 $\pm$ 1 \\
"                       & 1.6 & 12.6 $\pm$ 1 \\
"                       & 2.2 &  9.6 $\pm$ 1 \\
\enddata
\tablenotetext{a}{From NED.}
\tablenotetext{b}{From \citet{thuan96}, with colors defined 
within a 9\arcsec\ aperture centered on the nucleus.}
\tablenotetext{c}{In an aperture of 20\arcsec\ diameter.}
\end{deluxetable}

\clearpage

   \begin{deluxetable}{lrrr}
   \tablecolumns{4}
   \tablewidth{0pc}
   \tablecaption{Optical emission line intensities\tablenotemark{a}\label{tabint}}
   \tablehead{
   \colhead{Line} &
   \colhead{100$\times$$F(\lambda)/F({\rm H}\beta)$} &
   \colhead{100$\times$$I(\lambda)/I({\rm H}\beta)$} &
   \colhead{EW(H$\beta$)} 
   }
   \startdata
3697 H17                  &  0.5 $\pm$ 0.1&  0.7 $\pm$ 0.1&  0.4 \\
3704 H16                  &  1.1 $\pm$ 0.1&  1.6 $\pm$ 0.1&  0.9 \\
3712 H15                  &  0.8 $\pm$ 0.1&  1.2 $\pm$ 0.3&  0.7 \\
3727 [O {\sc ii}]         & 99.1 $\pm$ 1.4&144.4 $\pm$ 1.8& 86.4 \\
3750 H12                  &  2.0 $\pm$ 0.1&  2.9 $\pm$ 0.3&  1.8 \\
3771 H11                  &  2.5 $\pm$ 0.1&  3.5 $\pm$ 0.2&  2.2 \\
3797 H10                  &  3.3 $\pm$ 0.1&  4.7 $\pm$ 0.2&  2.9 \\
3820 He {\sc i}           &  1.7 $\pm$ 0.1&  2.4 $\pm$ 0.1&  1.5 \\
3835 H9                   &  5.3 $\pm$ 0.1&  7.3 $\pm$ 0.2&  4.7 \\
3868 [Ne {\sc iii}]       & 44.7 $\pm$ 0.7& 61.7 $\pm$ 0.8& 38.1 \\
3889 He {\sc i}+H8        & 13.8 $\pm$ 0.2& 18.9 $\pm$ 0.3& 11.9 \\
3968 [Ne {\sc iii}]+H7    & 24.7 $\pm$ 0.4& 32.9 $\pm$ 0.5& 21.4 \\
4009 He {\sc i}           &  0.4 $\pm$ 0.1&  0.5 $\pm$ 0.1&  0.3 \\
4026 He {\sc i}           &  2.4 $\pm$ 0.1&  3.1 $\pm$ 0.1&  2.0 \\
4038 N {\sc ii}+N {\sc ii}&  0.2 $\pm$ 0.1&  0.2 $\pm$ 0.1&  0.2 \\
4068 [S {\sc ii}]         &  1.4 $\pm$ 0.1&  1.8 $\pm$ 0.1&  1.2 \\
4076 [S {\sc ii}]         &  0.3 $\pm$ 0.1&  0.3 $\pm$ 0.1&  0.2 \\
4102 H$\delta$            & 21.8 $\pm$ 0.3& 27.8 $\pm$ 0.4& 19.2 \\
4121 He {\sc i}           &  0.4 $\pm$ 0.1&  0.4 $\pm$ 0.1&  0.3 \\
4143 He {\sc i}           &  1.1 $\pm$ 0.1&  1.4 $\pm$ 0.1&  1.0 \\
4181 N {\sc ii}           &  0.3 $\pm$ 0.1&  0.4 $\pm$ 0.1&  0.3 \\
4207 ?                    &  0.7 $\pm$ 0.1&  0.9 $\pm$ 0.1&  0.6 \\
4232 O {\sc i}            &  0.5 $\pm$ 0.1&  0.6 $\pm$ 0.1&  0.4 \\
4244 [Fe {\sc ii}]        &  0.7 $\pm$ 0.1&  0.9 $\pm$ 0.1&  0.7 \\
4267 C {\sc ii}           &  0.4 $\pm$ 0.1&  0.6 $\pm$ 0.1&  0.4 \\
4276 [Fe {\sc ii}]           &  0.5 $\pm$ 0.1&  0.6 $\pm$ 0.1&  0.5 \\ 
4287 [Fe {\sc ii}]           &  0.7 $\pm$ 0.1&  0.9 $\pm$ 0.1&  0.7 \\
4318 O {\sc ii}              &  0.5 $\pm$ 0.1&  0.6 $\pm$ 0.1&  0.5 \\
4340 H$\gamma$               & 42.7 $\pm$ 0.6& 50.2 $\pm$ 0.7& 40.6 \\
4363 [O {\sc iii}]           & 14.3 $\pm$ 0.2& 16.7 $\pm$ 0.3& 13.6 \\
4388 He {\sc i}              &  0.6 $\pm$ 0.1&  0.7 $\pm$ 0.1&  0.6 \\
4415 [Fe {\sc ii}]           &  1.0 $\pm$ 0.1&  1.2 $\pm$ 0.1&  1.0 \\
4452 [Fe {\sc ii}]           &  0.5 $\pm$ 0.1&  0.6 $\pm$ 0.1&  0.5 \\
4471 He {\sc i}              &  6.1 $\pm$ 0.1&  6.8 $\pm$ 0.1&  5.9 \\
4520 Si {\sc ii}             &  0.4 $\pm$ 0.1&  0.5 $\pm$ 0.1&  0.4 \\
4658 [Fe {\sc iii}]          &  2.8 $\pm$ 0.1&  3.0 $\pm$ 0.1&  2.6 \\
4702 [Fe {\sc iii}]          &  0.6 $\pm$ 0.1&  0.7 $\pm$ 0.1&  0.6 \\
4713 [Ar {\sc iv}]+He {\sc i}&  1.2 $\pm$ 0.1&  1.2 $\pm$ 0.1&  1.1 \\
4734 [Fe {\sc ii}]           &  0.6 $\pm$ 0.1&  0.6 $\pm$ 0.1&  0.6 \\
4755 [Fe {\sc iii}]          &  0.5 $\pm$ 0.1&  0.5 $\pm$ 0.1&  0.5 \\
4770 [Fe {\sc iii}]          &  0.2 $\pm$ 0.1&  0.2 $\pm$ 0.1&  0.2 \\
4778 [Fe {\sc iii}]          &  0.3 $\pm$ 0.1&  0.3 $\pm$ 0.1&  0.3 \\
4815 [Fe {\sc ii}]           &  0.4 $\pm$ 0.1&  0.4 $\pm$ 0.1&  0.4 \\
4861 H$\beta$                &100.0 $\pm$ 1.5&100.0 $\pm$ 1.5&106.9 \\
4881 [Fe {\sc ii}]           &  0.4 $\pm$ 0.1&  0.4 $\pm$ 0.1&  0.4 \\
4903 [Fe {\sc ii}]+Si {\sc ii}&  2.0 $\pm$ 0.1&  2.0 $\pm$ 0.1&  2.2 \\
4922 He {\sc i}              &  2.2 $\pm$ 0.1&  2.2 $\pm$ 0.1&  2.4 \\
4959 [O {\sc iii}]           &100.9 $\pm$ 1.5& 98.2 $\pm$ 1.4&110.3 \\
4987 [Fe {\sc iii}]          &  0.5 $\pm$ 0.1&  0.5 $\pm$ 0.1&  0.6 \\
5007 [O {\sc iii}]           &300.0 $\pm$ 4.3&287.9 $\pm$ 4.2&328.4 \\
\enddata
\tablenotetext{a}{$C$(H$\beta$) = 0.53, 
$F$(H$\beta$) = 7.5$\times$10$^{-17}$ W m$^{-2}$,
$I$(H$\beta$) = 25.4$\times$10$^{-17}$ W m$^{-2}$, 
EW$_{abs}$ = 0 \AA.} 
\end{deluxetable}

\clearpage

\begin{deluxetable}{rccl}
\footnotesize
\tablecaption{PAH Parameters Obtained by Fitting Lorentzian Profiles\label{tab:pahs}}
\tablewidth{0pt}
\tablehead{
\multicolumn{1}{c}{Wavelength\tablenotemark{a}} &
\multicolumn{1}{c}{Integrated\tablenotemark{b}} &
\multicolumn{1}{c}{Equivalent\tablenotemark{b}} &
\multicolumn{1}{c}{FWHM\tablenotemark{b}} \\
& \multicolumn{1}{c}{Flux} &
\multicolumn{1}{c}{Width} \\
\colhead{(\micron)}&\colhead{($10^{-17}$\,W\,m$^{-2}$)}
&\colhead{(\micron)}&\colhead{(\micron)}}
\startdata
5.759 	&  ~~3.3 (0.1) & 0.245 (0.029) & 0.044 (0.002) \\
6.225 	&  ~14.8 (0.4) & 1.030 (0.046) & 0.172 (0.005) \\
7.771 	&  ~54.8 (2.6) & 3.640 (0.427) & 0.837 (0.025)  \\
8.664   &  ~~6.1 (0.4) & 0.415 (0.064) & 0.200 (0.008)  \\
11.278	&  ~21.9 (1.0) & 1.360 (1.011) & 0.450 (0.260) \\
\enddata
\tablenotetext{a}{Rest wavelength, corrected 
for $z=0.00541$.}
\tablenotetext{b}{Standard deviations of the repeated measurements
 are given in parentheses.
The true uncertainty including calibration is probably $\sim$20\%.}
\end{deluxetable}

\clearpage

\begin{deluxetable}{crccccc}
\footnotesize
\tablecaption{Fine Structure Line Parameters 
Obtained by Fitting Gaussian Profiles\label{tab:lines}}
\tablewidth{0pt}
\tablehead{
\multicolumn{1}{c}{Line\tablenotemark{a}} &
\multicolumn{1}{c}{E$_{\rm ion}$\tablenotemark{b}} &
\multicolumn{1}{c}{Nominal} &
\multicolumn{1}{c}{Fitted\tablenotemark{c}} &
\multicolumn{1}{c}{Integrated\tablenotemark{d}} &
\multicolumn{1}{c}{Equivalent\tablenotemark{d}} &
\multicolumn{1}{c}{FWHM\tablenotemark{d}} \\
&& \multicolumn{1}{c}{Wavelength} &\multicolumn{1}{c}{Wavelength} 
& \multicolumn{1}{c}{Flux} & \multicolumn{1}{c}{Width} \\
&\colhead{(eV)} & \colhead{(\micron)} & \colhead{(\micron)} & 
\colhead{($10^{-17}$\,W\,m$^{-2}$)}
&\colhead{(\micron)} & \colhead{(\micron)}}
\startdata
\siv   & 34.8 & 10.511  & 10.512 &  ~~3.4 (0.15) & 0.27 (0.14) & ~0.026 (0.0004) \\
\neii  & 21.6 & 12.814 	& 12.814 &  ~~3.8 (0.3) & 0.14 (0.04) & 0.082 (0.062) \\
\neiii & 41.0 & 15.555	& 15.556 &  10.77 (0.5) & 0.22 (0.01) & 0.035 (0.005) \\
\siii  & 23.3 & 18.713	& 18.714 &  ~~7.4 (0.5) & 0.13 (0.01) & ~0.039 (0.0006) \\
\oiv   & 54.9 & 25.890	& 25.901 &  ~~1.2 (0.3) & 0.02 (0.01) & 0.026 (0.005) \\
\siii  & 23.3 & 33.481	& 33.480 &  10.85 (4.8) & 0.24 (0.10) & 0.055 (0.035) \\
\enddata
\tablenotetext{a}{Because \ariii\ and \arii\ are seen only in the low-resolution
spectra, we do not give their parameters here.}
\tablenotetext{b}{Ionization potential.}
\tablenotetext{c}{Rest wavelength, corrected for $z=0.00541$.}
\tablenotetext{d}{Standard deviations of repeated measurements
 are given in parentheses.
The true uncertainty including calibration is probably $\sim$20\%.}
\end{deluxetable}

\clearpage

   \begin{deluxetable}{lrrcrr}
   \tablecolumns{6}
   \tablewidth{0pc}
   \tablecaption{Parameters of narrow and broad emission lines\tablenotemark{a}
 \label{tabdeconv}}
   \tablehead{
\colhead{} & \multicolumn{2}{c}{Narrow component}&& 
\multicolumn{2}{c}{Broad component} \\ \cline{2-3} \cline{5-6}
   \colhead{Line} &
   \colhead{100$\times$$F(\lambda)/F({\rm H}\beta)$} &
   \colhead{FWHM (\AA)} &&
   \colhead{100$\times$$F(\lambda)/F({\rm H}\beta)$} &
   \colhead{FWHM (\AA)}
   }
   \startdata
3868 [Ne {\sc iii}]    & 38.1 & 3.5 && 52.5 & 8.5 \\
3889 He {\sc i}+H8     & 16.6 & 2.6 && 11.8 & 6.1 \\
4102 H$\delta$         & 23.8 & 2.6 && 20.7 & 6.7 \\
4340 H$\gamma$         & 42.2 & 2.6 && 42.5 & 6.7 \\
4363 [O {\sc iii}]     & \nodata & \nodata && 31.2 & 7.6 \\
4471 He {\sc i}        &  4.3 & 2.8 &&  8.6 & 8.4 \\
4686 He {\sc ii}       & \nodata & \nodata && 11.4 &31.7 \\
4861 H$\beta$          &100.0 & 2.6 &&100.0 & 7.6 \\
4959 [O {\sc iii}]     &126.0 & 2.7 && 69.2 & 6.6 \\
5007 [O {\sc iii}]     &399.6 & 2.9 &&175.9 & 7.5 \\
   \enddata
\tablenotetext{a}{
$F$(H$\beta$)$_{nar}$ = 4.0$\times$10$^{-17}$ W m$^{-2}$, 
$F$(H$\beta$)$_{br}$ = 3.5$\times$10$^{-17}$ W m$^{-2}$.} 
\end{deluxetable}

\clearpage

   \begin{deluxetable}{lrrrrr}
   \tablecolumns{6}
   \tablewidth{0pc}
   \tablecaption{Comparison of observed and modeled optical emission line
fluxes\tablenotemark{a} \label{tabcomparopt}}
   \tablehead{
   \colhead{} & \colhead{} & \multicolumn{2}{c}{CLOUDY Models\tablenotemark{b}} \\ \cline{3-4}
   \colhead{Line} &
   \colhead{Observ.} & \colhead{Model I} & \colhead{Model Ia\tablenotemark{c}} 
& \colhead{Model I/Obs.} & \colhead{Model Ia/Obs.}
   }
   \startdata
3727 [O {\sc ii}]    & 36.7& 73.3& 33.8& 2.00& 0.92 \\
3868 [Ne {\sc iii}]  & 15.7& 15.0& 14.5& 0.95& 0.93 \\
4363 [O {\sc iii}]   &  4.3&  5.1&  5.1& 1.18& 1.18 \\
4861 H$\beta$        & 25.4& 41.8& 28.4& 1.65& 1.11 \\
5007 [O {\sc iii}]   & 73.1& 64.7& 64.2& 0.89& 0.88 \\
   \enddata
\tablenotetext{a}{Fluxes are in units of 10$^{-17}$ W m$^{-2}$.}
\tablenotetext{b}{Fluxes are calculated assuming the distance $D$ = 21.7 Mpc.}
\tablenotetext{c}{Model Ia is the same as Model I except that the 
radius of the modeled H {\sc ii} region is 410 pc (log $r$ = 21.1) 
instead of 580 pc.}
\end{deluxetable}

\clearpage

   \begin{deluxetable}{lrrrrrrrrrrr}
   \tablecolumns{12}
   \tablewidth{0pc}
   \tablecaption{Comparison of observed and modeled MIR emission line
fluxes\tablenotemark{a} \label{tabcompar}}
   \tablehead{
   \colhead{} & \colhead{} & \multicolumn{5}{c}{CLOUDY Models\tablenotemark{b}} && \multicolumn{4}{c}{Model/Observ.} \\ \cline{3-7} \cline{9-12}
   \colhead{Line} &
   \colhead{Obs.} & \colhead{Mod.I} & \colhead{Mod.II} 
& \colhead{AGN} & \colhead{WNE-w}& \colhead{Shock\tablenotemark{c}} && \colhead{A\tablenotemark{d}}&\colhead{B\tablenotemark{e}}
&\colhead{C\tablenotemark{f}}&\colhead{D\tablenotemark{g}}
   }
   \startdata
10.5 [S {\sc iv}]    & 3.42& 0.19& 2.49& 0.65& 0.80&0.62&& 0.78& 0.97& 1.02& 0.96 \\
12.8 [Ne {\sc ii}]   & 3.81& 4.21& 0.54& 0.39& 0.03&0.05&& 1.25& 1.35& 1.25& 1.26 \\
15.6 [Ne {\sc iii}]  &10.77& 3.32& 6.88& 1.08& 1.02&0.88&& 0.95& 1.05& 1.04& 1.03 \\
18.7 [S {\sc iii}]   & 7.39& 2.90& 3.66& 0.66& 0.38&0.39&& 0.89& 0.98& 0.94& 0.94 \\
24.3 [Ne {\sc v}]    & ... &    0&    0& 0.34& 0.12&0.18&&  ...&  ...&  ...&  ... \\
25.9 [O {\sc iv}]    & 1.19&    0&    0& 1.12& 1.05&0.93&&    0& 0.94& 0.88& 0.78 \\
33.5 [S {\sc iii}]   &10.85& 3.33& 5.21& 0.88& 0.55&0.56&& 0.79& 0.87& 0.84& 0.84 \\
   \enddata
\tablenotetext{a}{Fluxes are in units of 10$^{-17}$ W m$^{-2}$.}
\tablenotetext{b}{Fluxes are calculated assuming the distance $D$ = 21.7 Mpc.}
\tablenotetext{c}{Shock velocity is 250 km s$^{-1}$.}
\tablenotetext{d}{A=I+II.}
\tablenotetext{e}{B=I+II+AGN.}
\tablenotetext{f}{C=I+II+WNE-w.}
\tablenotetext{g}{D=I+II+Shock.}
\end{deluxetable}



\begin{thebibliography}{}

\bibitem[Alonso-Herrero et al.(2001)]{alonso01} Alonso-Herrero, A., 
Quillen, A. C., Simpson, C., Efstathiou, A., \& Ward, M. J. 2001,
\aj, 121, 1369


\bibitem[Asplund et al.(2005)]{asplund05} Asplund, M., Grevesse, 
N., \& Sauval, A.~J.\ 2005, ASP Conf.~Ser.~336: Cosmic Abundances as 
Records of Stellar Evolution and Nucleosynthesis, 336, 25 


\bibitem[Beirao et al.(2006)]{b06} Beirao, P., Brandl, B. R., 
Devost, D., Smith, J. D., Hao, L., \& Houck, J. R. \apj, 643, L1 

\bibitem[Brandl et al.(2004)]{brandl04} Brandl, B.~R., et al.\ 
2004, \apjs, 154, 188 

\bibitem[Buchanan et al.(2006)]{buchanan06}  Buchanan, C. L., Gallimore, J. F., O'Dea, C. P., Baum, S. A., Axon, D. J., Robinson, A., Elitzur, M., \& 
Elvis, M. 2006, \aj, 132, 401 


\bibitem[Cardelli et al.(1989)]{cardelli89} Cardelli, J.~A., 
Clayton, G.~C., \& Mathis, J.~S.\ 1989, \apj, 345, 245 

\bibitem[Crowther et al.(1999)]{crowther99}  Crowther, P. A., Beck, S. C.,
Willis, A. J., Conti, P. S., Morris, P. W., \& Sutherland, R. S. 1999,
MNRAS, 304, 654


\bibitem[de Jong(1996)]{dejong96} de Jong, R.~S.\ 1996, \aap, 
313, 377 

\bibitem[De Robertis \& Osterbrock(1984)]{derobertis84} De Robertis, 
M.~M., \& Osterbrock, D.~E.\ 1984, \apj, 286, 171 

\bibitem[Dopita \& Sutherland(1996)]{dopita96}
 Dopita, M. A., \& Sutherland, R. S. 1996, \apjs, 102, 161

\bibitem[Draine(2006)]{draine06} Draine, B.~T. 2006, private communication

\bibitem[Draine \& Li(2001)]{draine01} Draine, B.~T., \& Li, A.\ 
2001, \apj, 551, 807 1


\bibitem[Enya et al.(2002)]{enya02} Enya, K., Yoshii, Y., Kobayashi, Y., 
Minezaki, T., Suganuma, M., Tomita, H., \& Peterson, B. A.,
2002, \apjs, 141, 23

\bibitem[Fazio et al.(2004)]{fazioirac} Fazio, G.~G., et al.\ 
2004, \apjs, 154, 10 

\bibitem[Ferland(1996)]{cloudy} Ferland, G. J. 1996, Hazy: A brief Introduction to CLOUDY 
(Univ. Kentucky Dept. Phys. Astron. Internal Rep.)

\bibitem[Ferland et al.(1998)]{ferland98} Ferland, G. J., Korista, K. T., Verner, D. A., Ferguson, J. W., 
Kingdon, J. B., \& Verner, E. M. 1998, \pasp, 110, 761


\bibitem[Gil de Paz et al.(2003)]{gildepaz03} Gil de Paz, A., 
Madore, B.~F., \& Pevunova, O.\ 2003, \apjs, 147, 29 



\bibitem[Guseva, Izotov \& Thuan (2000)]{guseva00}
Guseva, N. G., Izotov, Y. I., \& Thuan, T. X. 2000, \apj, 531, 776

\bibitem[Guseva, Izotov \& Thuan (2006)]{guseva06}
Guseva, N. G., Izotov, Y. I., \& Thuan, T. X. 2006, \apj, 644, 890


\bibitem[Hawarden et al.(2001)]{hawarden01} Hawarden, T. G., Leggett, S. K., 
Letawsky, M. B., Ballantyne, D. R., \& Casali, M. M. 2001, \mnras, 325, 563 

\bibitem[Helou et al.(2000)]{helou00} Helou, G., Lu, N. Y., 
Werner, M. W., Malhotra, S., \& Silbermann, N. 2000, \apj, 532, L21 

 

\bibitem[Houck et al.(2004)]{houckirs} Houck, J.~R., et al.\ 
2004, \apjs, 154, 18 



\bibitem[Hunt \& Giovanardi(1992)]{hg92}
Hunt, L.K., \& Giovanardi, C. 1992, \aj, 104, 1018

\bibitem[Hunt et al.(2002)]{hunt02} Hunt, L.~K., Giovanardi, 
C., \& Helou, G.\ 2002, \aap, 394, 873 

\bibitem[Hunt et al.(1997)]{hunt97} Hunt, L.~K., Malkan, 
M.~A., Salvati, M., Mandolesi, N., Palazzi, E., \& Wade, R.\ 1997, \apjs, 
108, 229 

\bibitem[Hunt, Thuan, \& Izotov (2003)]{hunt03}
Hunt, L.K., Thuan, T. X., \& Izotov, Y. I. 2003, \apj, 588, 281

\bibitem[Hunt et al.(2005)]{hunt05} Hunt, L. K., Bianchi, S., \& 
Maiolino, R.\ 2005, \aap, 434, 849 

\bibitem[Hunt et al.(2006)]{hunt06} 
Hunt, L. K., Thuan, T. X., Sauvage, M., \& Izotov, Y. I. 2006, 
\apj, 653, 222 



 
\bibitem[Ivezi\'c \& Elitzur(1997)]{elitzur} Ivezi\'c, Z.~\& 
Elitzur, M.\ 1997, MNRAS, 287, 799 

\bibitem[Izotov et al. (1992)]{izotov92} 
Izotov, Y. I., Lipovetsky, V. A., Guseva, N. G., \& Kniazev, A. Y. 1992,
in The Feedback of Chemical Evolution on the Stellar Content of 
Galaxies, ed. D. Alloin \& G. Stasi\'nska (Paris: Paris Obs. Publ.), p. 138

\bibitem[Izotov, Thuan \& Lipovetsky (1994)]{itl94}
 Izotov, Y. I., Thuan, T. X., \& Lipovetsky, V. A. 1994, \apj, 435, 647 

\bibitem[Izotov et al. (2001)]{izotov01} Izotov, Y. I., Chaffee, F. H.,
\& Schaerer, D. 2001, \aap, 378, L45

\bibitem[Izotov et al. (2004)]{izotov04} Izotov, Y. I., Noeske, K. G., 
Guseva, N. G., Papaderos, P., Thuan, T. X., \& Fricke, K. J. 
2004, \aap, 415, L27

\bibitem[Johnson et al.(2004)]{johnson04} Johnson, K.~E., 
Indebetouw, R., Watson, C., \& Kobulnicky, H.~A.\ 2004, \aj, 128, 610 

\bibitem[Lebouteiller et al.(2007)]{l07} Lebouteiller, V., Brandl, B., 
Bernard-Salas, J., Devost, D., \& Houck, J. R. 2007, \apj, 665, 390 

\bibitem[Leitherer et al.(1999)]{sb99} Leitherer, C., et al.\ 1999, \apjs, 123, 3 

\bibitem[Li \& Draine(2001)]{li01} Li, A., \& Draine, B.~T.\ 
2001, \apj, 554, 778 


\bibitem[Loose \& Thuan(1985)]{loose85} Loose, H.-H. \& Thuan, T. X. 1985, 
in Star-Forming Dwarf Galaxies 
and related objects, eds. D. Kunth, T.X. Thuan \& J.T.T. Van (Gif-sur-Yvette:
Editions Frontieres), p. 73 


\bibitem[Makovoz \& Marleau(2005)]{mopex} Makovoz, D., \& 
Marleau, F.~R.\ 2005, \pasp, 117, 1113 

\bibitem[Mart\'in-Hern\'andez et al. (2002)]{m02} Mart\'in-Hern\'andez, N. L.,
et al. 2002, \aap, 381, 606

\bibitem[Melnick et al.(2008)]{melnick08} Melnick, G.~J., Tolls, 
V., Neufeld, D.~A., Yuan, Y., Sonnentrucker, P., Watson, D.~M., Bergin, 
E.~A., \& Kaufman, M.~J.\ 2008, ArXiv e-prints, 805, arXiv:0805.0573 


\bibitem[Mouhcine \& Lan{\c c}on(2002)]{mouhcine02} Mouhcine, M., 
\& Lan{\c c}on, A.\ 2002, \aap, 393, 149 






\bibitem[Origlia et al.(1999)]{origlia99} Origlia, L., Goldader, 
J.~D., Leitherer, C., Schaerer, D., \& Oliva, E.\ 1999, \apj, 514, 96 

\bibitem[Peletier \& Balcells(1997)]{peletier97} Peletier, R.~F., 
\& Balcells, M.\ 1997, New Astronomy, 1, 349 



\bibitem[Reach et al.(2005)]{reach05} Reach, W.~T., et al.\ 
2005, \pasp, 117, 978 

\bibitem[Rieke et al.(2004)]{riekemips} Rieke, G.~H., et al.\ 
2004, \apjs, 154, 25 


\bibitem[Schaerer \& de Koter (1997)]{schaerer97} 
Schaerer, D., \& de Koter, A. 1997,\aap, 322, 598





\bibitem[Schlegel et al.(1998)]{schlegel98} Schlegel, D.~J., 
Finkbeiner, D.~P., \& Davis, M.\ 1998, \apj, 500, 525 

\bibitem[Silva et al.(1998)]{silva98} Silva, L., Granato, 
G.~L., Bressan, A., \& Danese, L.\ 1998, \apj, 509, 103 

\bibitem[Tantalo et al. (1996)]{tantalo96} Tantalo, R., Chiosi, C., Bressan, 
A., \& Fagotto, F. 1996, \aap, 311, 361	


\bibitem[Thuan \& Izotov(2005)]{thuanizotov05} Thuan, T. X., \& Izotov, Y. I. 2005, \apjs, 161, 240

\bibitem[Thuan et al.(1996)]{thuan96} Thuan, T.~X., Izotov, 
Y.~I., \& Lipovetsky, V.~A.\ 1996, \apj, 463, 120 


\bibitem[Thuan et al.(1999)]{thuanhi}
Thuan, T. X., Lipovetsky, V. A., Martin, J.-M., \& Pustilnik, S. A. 1999, \aaps,
139, 1



\bibitem[V{\'a}zquez \& Leitherer(2005)]{vazquez05} V{\'a}zquez, 
G.~A., \& Leitherer, C.\ 2005, \apj, 621, 695 

\bibitem[Verma et al.(2003)]{verma03} Verma, A., Lutz, D., 
Sturm, E., Sternberg, A., Genzel, R., \& Vacca, W.\ 2003, \aap, 403, 829 


\bibitem[Weedman et al.(2005)]{weedman05}
Weedman, D.~W., et al. 2005, \apj, 633, 706 

\bibitem[Werner et al.(2004)]{werner04} Werner, M.~W., et al.\ 
2004, \apjs, 154, 1 

\bibitem[Whitford(1958)]{w58}
Whitford, A. E. 1958, \aj, 63, 201


\bibitem[Whittle(1985)]{whittle85} Whittle, M.\ 1985, \mnras, 
216, 817 

\bibitem[Wu et al.(2006)]{wu06} Wu, Y., Charmandaris, V. Hao, L., Brandl, B. R.,
Bernard-Salas, J., Spoon, H. W. W., \& Houck, J. R. 2006, \apj, 639, 157


\end{thebibliography}
\end{document}